\documentstyle[12pt]{article}
\begin{document}

\def\la{\mathrel{\mathpalette\fun <}}
\def\ga{\mathrel{\mathpalette\fun >}}
\def\fun#1#2{\lower3.6pt\vbox{\baselineskip0pt\lineskip.9pt
  \ialign{$\mathsurround=0pt#1\hfil##\hfil$\crcr#2\crcr\sim\crcr}}}
%less than approximately and greater than approximately
\newcommand{\dd}{\mbox{d}}
\newcommand{\vecc}[1]{\mbox{\boldmath $#1$}}
\begin{titlepage}
%\rightline{CERN-TH/95-313}
%\rightline{UPRF-95-438}
%\vskip 20.0pt
\centerline{\Large \bf Small--Angle Electron--Positron Scattering}
\vskip 10.0pt
\centerline{\Large \bf with a Per Mille Accuracy\footnote{Work supported by the
Istituto Nazionale di Fisica Nucleare (INFN). INTAS Grant 1867--93. }}
\vskip 1.0cm
\centerline{A.B.~Arbuzov~$^{a}$, V.S.~Fadin~$^{b}$,
E.A.~Kuraev~$^{a}$, L.N.~Lipatov~$^{c}$,
N.P.~Merenkov~$^{d}$}
\vskip 0.3cm
\centerline{and}
\vskip 0.3cm
\centerline{L.~Trentadue~$^{e}$\footnote{On leave from Dipartimento
di Fisica, Universit\'a di Parma, Parma, Italy.}\footnote{INFN, Gruppo Collegato
di Parma, Parma, Italy.}}
\vskip 1.0cm
\begin{center}
{$^a$ \it Joint Institute for Nuclear Research\\
Dubna, Moscow region, 141980, Russia \\
\vskip.3cm
$^b$ \it Budker Institute for Nuclear Physics \\ Novosibirsk State University,
630090, Novosibirsk, Russia \\
\vskip.3cm
$^c$ \it St.-Petersburg Institute of Nuclear Physics\\
Gatchina, Leningrad region, 188350, Russia \\
\vskip.3cm
$^d$ \it Institute of Physics and Technology,
Kharkov, 310108, Ukraine \\
\vskip.3cm
$^e$ \it Theoretical Physics Division, CERN, \\ CH--1211 Geneva 23, Switzerland}
\end{center}
\vskip 10.0pt
\begin{abstract}
The elastic and inelastic high--energy small--angle electron--positron
scattering is considered. All radiative corrections to the cross--section
with the relative accuracy $\delta\sigma/ \sigma = 0.1 \% $
are explicitly taken into account. According to the generalized eikonal
representation for the elastic amplitude,
in higher orders only diagrams with one exchanged photon
may be considered. Single photon emission with  radiative
corrections and next--to--leading two--photon and pair production
diagrams are evaluated, together with leading three--loop
corrections. All contributions have been calculated analytically.
We integrate the calculated distributions over typical for
LEP~1 experiments intervals of angles and energies.
To the leading approximation, the results are shown to be
described in terms of kernels of electron structure functions.
Some numerical results are presented.
\end{abstract}
\vskip 1cm
PACS numbers 12.15.Lk,  12.20.-m,  12.20.Ds, 13.40.-f
%\hrule
\vskip.7cm
\noindent
%\vfill
%\leftline{CERN-TH/95-313}
%\leftline{November 1995}
\end{titlepage}
%\eject
\textheight 210mm
\topmargin 2mm
\newpage

\section{Introduction}

An accurate verification of the Standard
Model is one of the primary aims of LEP \cite{r1}.
While electroweak radiative corrections to the
$s$-channel annihilation process and to large--angle Bhabha scattering
allow a direct extraction of the Standard Model parameters, small
angle Bhabha cross--section affects, as an overall
normalization condition, all observable cross--sections and
represents an equally unavoidable condition toward a precise
determination of the Standard Model parameters.
 The small--angle Bhabha scattering process is used
to measure the luminosity of electron--positron colliders. At LEP an
experimental accuracy on the luminosity of
\begin{eqnarray}\label{eq1}
      |\frac{\delta\sigma}{ \sigma}|< 0.001
\end{eqnarray}
has been reached \cite{r1a}. However, to obtain the total accuracy,
a systematic theoretical error must also be added. This precision calls for
an equally accurate theoretical expression for the Bhabha scattering
cross--section in order to extract the Standard Model
parameters from the observed distributions. An accurate determination of
the small--angle Bhabha cross--section and of the luminosity directly affects
the determination of absolute cross--sections such as, for example,
the determination of the invisible width and of the number of massless
neutrino species $N_{\nu}$ \cite{r111}.

In recent years a considerable attention has been
devoted to the Bhabha process \cite{r2,r3,r4}. The
reached accuracy is however still inadequate \cite{r1a}. According to these
evaluations the theoretical estimates are still incomplete; moreover,
they are somewhat larger ($\sim$ a factor 2) than
the projected theoretical and experimental precision \cite{r1a} and
are comparable to the currently published experimental precision.

The process that  will be considered  in this work is that of Bhabha scattering
when electrons and positrons are emitted at small angles
with respect to the initial electron and positron directions. We have examined
the
radiative processes inclusively accompanying the main
e$^+$e$^-$$\rightarrow$ e$^+$e$^-$ reaction at
high energies, when both the scattered electron and positron are tagged
within the counter aperture.

We assume that the center-of-mass energies are within the range of the LEP
collider $2\epsilon=\sqrt{s}=90$ -- $200\;$GeV
and the scattering angles are within the range $\theta \simeq 10$ --
$150\;$mrad.
We assume that the charged-particle detectors have the following polar angle
cuts:

\begin{eqnarray}\label{eq2}
\theta_1 < \theta_-=\widehat{\vecc{p}_1 \vecc{q}_1\,} \equiv
\theta < \theta_3, \qquad
\theta_2 < \theta_+=\widehat{\vecc{p}_2 \vecc{q}_2\,}<\theta_4,
\qquad 0.01\la \theta_i \la 0.1 \;\mbox{rad}\;, \label{angles}
\end{eqnarray}
where $\vecc{p}_1 , \ \vecc{q}_1\,\; (\vecc{p}_2 , \ \vecc{q}_2\,)$ are
the momenta of the initial and of the scattered electron (positron) in the
center-of-mass frame.

In this paper we present the results of our calculations
of the electron--positron scattering cross--section with an accuracy of
${\cal O }(0.1\%)$.
The squared matrix elements of the various exclusive processes
inclusively contributing to the e$^+$e$^-$ $\rightarrow$ e$^+$e$^-$
reaction are integrated in order to define an experimentally measurable
cross--section according to suitable restrictions on the angles and energies
of the detected particles. The different contributions to the electron and
positron distributions, needed for the required accuracy, are presented
using analytical expressions.

In order to define the angular range of interest and the implications on
the required accuracy, let us first briefly discuss, in a general way, the
angle-dependent
corrections to the cross--section.

We consider  e$^+$e$^-$  scattering at angles as defined in Eq.~(\ref{angles}).
Within this region, if one expresses the cross--section by means of a series
expansion in terms of angles, the main contribution to the
cross--section
${\dd\sigma}/{\dd\theta^2}$
comes from the diagrams for the scattering amplitudes containing one  exchanged
 photon
in the $t$-channel. These diagrams, as well known, show a
singularity of the type $\theta^{-4}$ for $\theta \rightarrow 0$, e.g.
\begin{eqnarray}
 \frac{\dd\sigma}{\dd\theta^2} & \sim & \theta^{-4}\;\;\;.\nonumber
\end{eqnarray}

Let us now estimate the correction of order $\theta^2$ to this contribution.
If
\begin{eqnarray}\label{eq3}
 \frac{\dd\sigma}{\dd\theta^2} & \sim & \theta^{-4}(1+c_1\theta^2)\;\;\; ,
\end{eqnarray}
then, after integration over  $\theta^2$  in the angular range of
Eq.~(\ref{angles}),
we obtain:

\begin{eqnarray}\label{eq4}
\int \limits_{\theta^2_{\mbox{\tiny min}}}^{\theta^2_{\mbox{\tiny max}}}
\frac{\dd\sigma}{\dd\theta^2}\;\dd\theta^2
& \sim & \theta^{-2}_{\mbox{\tiny min}}(1+c_1\theta^2_{\mbox{\tiny min}}
\ln \frac{\theta^2_{\mbox{\tiny max}}}{\theta^2_{\mbox{\tiny min}}}).
\end{eqnarray}
 We see that, for  $\theta_{\mbox{\tiny min}} = 50 \;$mrad and
$\theta_{\mbox{\tiny max}} = 150\;$mrad (we have taken
the case where the $\theta^2$  corrections are maximal), the relative
contribution of the $\theta^2$ terms is about  $5 \times 10^{-3} c_1.$
Therefore,
the terms of relative order $\theta^2$  must be kept only in the
Born cross--section where the coefficient $c_1$ is not small.
In higher orders of the
perturbative expansion the coefficient
$c_1$  contains at least one factor ${\alpha}/{\pi}$ and therefore these
terms can be safely omitted. This implies that, within our accuracy,
only radiative corrections from the scattering-type diagrams contribute
\cite{rx}. Furthermore only diagrams with
one photon exchanged in the $t$-channel should be taken into account
according to the generalized eikonal representation see Eq.(16) below.

Having as a final goal for the experimental cross--section the relative
accuracy of Eq.~(\ref{eq1}), and taking into account that the minimal value of
the squared momentum transfer
$Q^2 = 2\epsilon^2 (1-\cos\theta)$ in the region defined in Eq.~(\ref{angles})
is of the
order of $1\;\mbox{GeV}^2$, we  may omit  in the following also the terms
appearing in the radiative corrections of the type
${m^2}/{Q^2}$, with $m$ equal to the electron mass.

The contents of this paper can be outlined as follows. In Section 2
we discuss the Born cross--section $\dd\sigma^B$ by taking the
$Z^0$-boson exchange into account and we compute
the corrections to it due to the virtual and  real soft-photon
emission. We define also an {\em experimentally measurable\/} cross--section
$\sigma_{\mbox{\tiny exp}}$ with the
experimental cuts on angles and energies taken into account and we discuss
how to obtain it from the differential distributions.
We present the results, as discussed above,
in the form of an expansion in terms of the scattering angle $\theta$.
The ratio
$\Sigma={\sigma_{\mbox{\tiny exp}}}/{\sigma_0}$ is introduced by  normalizing
$\sigma_{\mbox{\tiny exp}}$ with
 respect to the cross--section $\sigma_0={4 \pi \alpha^2}/
{\epsilon^2\theta^2_1}$.
In Section~3, by using a simplified, but still suitable to
obtain the required accuracy, version of the differential cross--section
for the small--angle scattering, we discuss the contribution to
$\sigma_{\mbox{\tiny exp}}$ from the single bremsstrahlung process.
The details of the Sudakov technique which we use to calculate the 
hard-photon emission are given in Appendix~A.
In Section~4  we find all  corrections of ${\cal O}(\alpha^2)$ to
$\sigma_{\mbox{\tiny exp}}$ caused by virtual and
real photons emission as well as pair production.
In Section~5 we consider the virtual and soft-photon emission
accompanying the single photon bremsstrahlung process. The details of this
derivation are given in Appendices~B and C.
In Section~6 we consider the double hard-photon emission process in both
the same-side and opposite-side cases. Details are given in Appendix~D.
In Section 7 we consider the hard pair production process in both the
collinear and semi--collinear kinematical region. The details of this
calculation are given in
Appendix F. In Appendix~D the
expressions for the leading logarithmic approximation in terms of structure
functions factorization are given. The details of the cancellation of the
$\Delta$-dependence are presented in Appendix~E, $\Delta$ is a small
auxiliary parameter $(\Delta\ll 1)$ which separate the processes
of soft and hard real photon emission (see eq.~(15)).
In Section 8 the expressions to leading logarithmic
${\cal O}(\alpha^3)$ for the e$^+$e$^-$ and e$^+$e$^-$$\gamma$ radiative
processes are obtained. In Section 9, finally, estimates of the neglected
terms together with numerical results are presented.

A less detailed derivation of these results has  been reported elsewhere
\cite{r33}.

\section{Born cross--section and
%\\
one-loop virtual and soft corrections}

The Born cross--section for Bhabha scattering within the Standard Model is
well known \cite{r4}:
\begin{equation} \label{eq5}
\frac{\dd\sigma^B }{\dd\Omega }\, = \,\frac{\alpha^2 }{8s} \bigl\{
4B_1 +(1-c)^2B_2 +(1+c)^2 B_3 \bigr\},
\end{equation}
where
\begin{eqnarray*}
B_1 &=& (\frac{s}{t})^2 \left|1+(g^2_v-g^2_a)\xi \right|^2  ,
\qquad B_2= \left| 1+(g^2_v-g^2_a) \chi \right|^2 ,       \\
B_3 &=&\frac{1}{2} \left| 1+\frac{s}{t}+(g_v+g_a)^2(\frac{s}{t}
\xi +\chi)\right|^2
+ \frac{1}{2} \left| 1+\frac{s}{t}+(g_v-g_a)^2(\frac{s}{t}
\xi +\chi)\right|^2, \\
\chi &=& \frac{\Lambda s}{s-m^2_z+iM_Z \Gamma _Z}\, , \qquad
\xi = \frac{\Lambda t}{t-M^2_Z}\, ,
\\ \Lambda &=& \frac{G_FM^2_Z}{2\sqrt 2 \pi \alpha}=(\sin 2\theta
_w)^{-2},\quad
g_a=-\frac{1}{2},\quad
g_v=-\frac{1}{2} (1-4 \sin^2\theta_w), \\
s &=& (p_1 +p_2)^2 =4\varepsilon^2, \qquad
t=-Q^2=(p_1-q_1)^2=-\frac{1}{2}\;s\;(1-c),\\
c &=& \cos\theta ,\qquad \theta =\widehat{\vecc{p}_1 \vecc{q}_1}.
\end{eqnarray*}
Here $\theta_w$ is the Weinberg angle.
In the small--angle limit
$(c=1 - \theta^2/2 + \theta^4/24 + \dots)$,
expanding formula (\ref{eq5}) leads to
\begin{eqnarray} \label{eq6}
\frac{\dd\sigma^B}{\theta\dd\theta}=\frac{8 \pi \alpha^2}{\varepsilon^2 \theta^4}
(1-\frac{\theta^2}{2}+\frac{9}{40} \theta^4+\delta_{\mbox{\tiny weak}}),
\end{eqnarray}
where $\varepsilon=\sqrt{s}/2$ is the electron or positron initial energy
and the weak correction term $\delta_{\mbox{\tiny weak}}$, connected with
diagrams with $Z^0$-boson
exchange, is given by the expression:
\begin{eqnarray} \label{eq7}
\delta_{\mbox{\tiny weak}}=2g_v^2 \xi-\frac{\theta^2}{4}(g_v^2+g_a^2)
\mbox{Re}\,\chi
+ \frac{\theta^4}{32}(g_v^4+g_a^4+6g_v^2g_a^2)|\chi|^2.
\end{eqnarray}

One can see from Eq.~(\ref{eq7}) that the contribution $c^w_1$  of the
weak correction $\delta_{\mbox{\tiny weak}}$  into the coefficient $c_1$ introduced
in Eq.~(\ref{eq3})
\begin{equation}
c^w_1 \la 2g_v^2+\frac{(g_v^2+g_a^2)}{4} \frac{M_Z}{\Gamma_Z}
+\theta^2_{\mbox{\tiny max}}\frac{(g_v^4+g_a^4+6g_v^2g_a^2)}{32}\frac{M^2_Z}{\Gamma^2_Z}
\simeq 1.
\end{equation}
The contribution connected with $Z^0$-boson exchange diagrams
does not exceed $0.3\%$ for typical energy and angle ranges.
Radiative corrections to $Z^0-\gamma$ interference contributions
were considered in detail in papers by W.~Beenakker and
B.~Pietrzyk~\cite{r2}. They are not small and should be
taken into account in an analysis of experimental data.
We will not touch the subject in this publication.

In the pure QED case one-loop radiative corrections to  Bhabha cross--section
were
calculated a long time ago \cite{r5}.
Taking into account a contribution from soft-photon emission with energy less
than a  given finite threshold $\Delta \varepsilon$, we have here for
the cross--section $d \sigma^{(1)}_{QED}$, in the one-loop approximation:
\begin{eqnarray} \label{eq9}
\frac{\dd\sigma^{(1)}_{QED}}{\dd c}=\frac {\dd\sigma^{B}_{QED}}{\dd c}\;
(1 + \delta_{\mbox{\tiny virt}} + \delta_{\mbox{\tiny soft}}),
\end{eqnarray}
where $\dd\sigma^{B}_{QED}$ is the Born cross--section in the pure QED case
(it is equal to $\dd\sigma^{B}$ with \\ $g_a =g_v = 0$) and
\begin{eqnarray}
\delta_{\mbox{\tiny virt}} + \delta_{\mbox{\tiny soft}}
&=& 2 \frac{\alpha}{\pi} \; \biggl[ 2
\biggl(1- \ln\frac{4\varepsilon^2}{m^2}
+ 2\ln(\cot\frac {\theta}{2})\biggr)
\ln\frac{\varepsilon}{\Delta\varepsilon}
+ \int\limits_{\cos^2(\theta/2)}^{\sin^2(\theta/2)}
\frac{\dd x}{x} \ln(1-x) \nonumber\\
&-& \frac{23}{9} + \frac{11}{6} \; \ln\frac{4\varepsilon^2}{m^2} \biggr]\;
+\; \frac{\alpha}{\pi} \;\frac{1}{(3+ c^2)^2}\;\biggl[
\frac{\pi^2}{3}\;(2 c^4 -3 c^3 - 15 c) \nonumber \\
&+& 2\;(2 c^4- 3 c^3 + 9 c^2+ 3 c+ 21)\;\ln^2(\sin\frac{\theta}{2})
\nonumber \\
&-& 4\;(c^4 + c^2 - 2 c)\;
\ln^2\cos\frac{\theta}{2} - 4\;( c^3 + 4 c^2+ 5 c+ 6)\;
\ln^2(\tan\frac{\theta}{2}) \nonumber \\
&+& \frac{2}{3}\;(11 c^3+33 c^2+21 c+111)\;\ln(\sin\frac{\theta}{2})
+2\;(c^3-3 c^2+7 c-5)\;\ln(\cos\frac{\theta}{2})
\nonumber \\ \nonumber
&+& 2\;(c^3+3 c^2+3 c+9) \;\delta_t-2\;( c^3+3 c)(1-c) \;\delta_s \biggr].
\end{eqnarray}
The value $\delta_t$  ($\delta_s$)   is defined by contributions to the photon
vacuum polarization function $\Pi(t)$  ($\Pi(s)$)  as follows:
\begin{eqnarray}
\Pi(t)=\frac{\alpha}{\pi} \;\biggl(\delta_t+\frac{1}{3}L-\frac{5}{9}\biggr)
+\frac{1}{4}\;(\frac{\alpha}{\pi})^2 L,
\end{eqnarray}
where
\begin{eqnarray}
L=\ln\frac{Q^2}{m^2},\qquad  Q^2=-t=2\varepsilon^2 (1-c),
\end{eqnarray}
and we took into account the leading part of the two--loop contribution
in the
polarization operator. In the Standard Model, $\delta_t$ contains
contributions of muons, tau-leptons, $W$-bosons and hadrons:
\begin{eqnarray}
\delta_t= \delta_t^{\mu} + \delta_t^{\tau} + \delta_t^{W}
+\delta_t^H, \qquad \delta_s=\delta_t\;(Q^2\rightarrow-s),
\end{eqnarray}
the first three contributions are
theoretically calculable and can be given as:
\begin{eqnarray}
&& \delta_t^{\mu}=\frac{1}{3}\ln\frac{Q^2}{m_{\mu}^2}-\frac{5}{9}, \nonumber\\
 && \delta_t^{\tau}= \frac{1}{2}\;v_{\tau}\;(1-\frac{1}{3}v^2_{\tau})
\;\ln\frac{v_{\tau}+1}{v_{\tau}-1}+\frac{1}{3}\;v^2_{\tau}
- \frac{8}{9}, \qquad
v_{\tau}=\sqrt{1+\frac{4m_{\tau}^2}{Q^2}}\, , \\
 && \delta_t^{W}= \frac{1}{4}\;v_{W}\;(v^2_{W}-4)
\;\ln\frac{v_{W}+1}{v_{W}-1}-\frac{1}{2}\;v^2_{W}
+\frac{11}{6}, \qquad
v_W=\sqrt{1+\frac{4M_{W}^2}{Q^2}}\, .   \nonumber
\end{eqnarray}
The contribution of hadrons cannot be calculated theoretically; instead,
it can be given as integration of the experimentally measurable
cross--section:
\begin{eqnarray}
&& \delta_t^H=\frac{Q^2}{4\pi\alpha^2}\int\limits_{4m_{\pi}^2}^{+\infty}
\frac{\sigma^{e^+e^-\rightarrow h}\;(x)}{x+Q^2}\;\dd x.
\end{eqnarray}
For numerical calculations we will use for $\Pi(t)$ the results
of Ref.~\cite{r6}.

In the small scattering angle limit
we can present (\ref{eq9}) in the following form:
\begin{eqnarray}
&& \frac{\dd\sigma^{(1)}_{QED}}{\dd c}=\frac {\dd\sigma^{B}_{QED}}{\dd c}\;
(1-\Pi(t))^{-2}\;(1+\delta),
\\ \nonumber &&
\delta=2 \frac{\alpha}{\pi}\;\biggl[2(1-L)\ln\frac{1}{\Delta}+\frac{3}{2}L
- 2\biggr] + \frac{\alpha}{\pi}\;\theta^2\;\Delta_{\theta}
+\frac{\alpha}{\pi}\;\theta^2\; \ln\Delta,
\\ \nonumber &&
\Delta_{\theta}=\frac{3}{16}l^2+\frac{7}{12}l-\frac{19}{18}
+\frac{1}{4}\;(\delta_t-\delta_s),
\\ \nonumber &&
\Delta=\frac{\Delta\varepsilon}{\varepsilon},  \qquad l=\ln\frac{Q^2}{s}\simeq
\ln \;\frac {\theta^2}{4}.
\end{eqnarray}
This representation gives us a possibility to verify explicitly that
the terms of relative order $\theta^2$ in the radiative corrections are
small.
Taking into account that the large contribution proportional to $\ln\Delta$
disappears when we add the cross--section for the hard emission, we can
verify again that such terms can be
neglected. Therefore we will omit in higher orders the annihilation diagrams and
multiple-photon exchange diagrams in the scattering channel.
The second simplification is justified by the generalized eikonal
representation for small--angle scattering amplitudes.
In particular, for the case of elastic processes we have  \cite{r7}:

\begin{eqnarray} \label{eq16}
A(s,t)=A_0(s,t)\;F_1^2(t)\;(1-\Pi(t))^{-1} \;\mbox{e}^{i\varphi(t)}\;
\left[1+{\cal {O}}\biggl(\frac{\alpha}{\pi}\;\frac{Q^2}{s}\biggr)\right],\quad
s\gg Q^2\gg m^2,
\end{eqnarray}
where $A_0(s,t)$ is the Born amplitude, $F_1(t)$ is the Dirac form
factor and $\varphi(t)=-\alpha\; \ln(Q^2/\lambda^2)$ is the Coulomb phase,
$\lambda$ is the {\em photon mass\/} auxiliary parameter.
The  eikonal representation is violated  at a three--loop level, but,
fortunately, the corresponding contribution to the
Bhabha cross--section is small enough $(\sim\alpha^5)$ and can be neglected
for our purposes.  We may consider the eikonal representation as correct
within the required  accuracy\footnote{
Result obtained in paper \cite{r8}, we believe, is incorrect.
It contradicts to the well established result of D.~Yennie et al. \cite{YFS}
about cancelation of infrared singularities.}.

Let us now introduce the dimensionless quantity
$\Sigma=Q_1^2\;\sigma_{\mbox{\tiny exp}}/(4\pi\alpha^2)$,
with $Q_1^2=\varepsilon^2
\theta_1^2$, where  $\sigma_{\mbox{\tiny exp}}$
is the Bhabha--process cross--section
integrated over the typical experimental energy and angular
ranges\footnote{Really this quantity corresponds to some
{\em ideal\/} detectors. It is intended for comparisons
with the results of Monte Carlo event generators.}:
\begin{eqnarray}
\Sigma=\frac{Q_1^2}{4\pi\alpha^2}\int\!\dd x_1 \int\!\dd x_2
\;\Theta(x_1x_2-x_c)\int\!\dd^2\vecc{q}_1^{\bot}\;\Theta_1^c
\int\!\dd^2\vecc{q}^{\bot}_2
\;\Theta_2^c\;\frac{\dd\sigma^{e^+e^-\rightarrow e^+(\vecc{q}^{\bot}_2,x_2)\,
e^-(\vecc{q}^{\bot}_1,x_1)+X}}{\dd x_1\dd^2\vecc{q}^{\bot}_1\dd x_2
\dd^2\vecc{q}^{\bot}_2}\, ,
\end{eqnarray}
where $x_{1,2}$, $\vecc{q}^{\bot}_{1,2}$  are the energy fractions and the
transverse components of the momenta of the electron and
positron in the final state, $sx_c$ is the experimental cut--off on their
invariant mass squared and the functions
$\Theta_i^c$ do take into account the angular cuts (\ref{eq2}):
\begin{eqnarray}
\Theta_1^c=\Theta(\theta_3-\frac{|\vecc{q}^{\bot}_1|}{x_1\varepsilon})\;\;
\Theta(\frac{|\vecc{q}^{\bot}_1|}{x_1\varepsilon}-\theta_1), \qquad
\Theta_2^c=\Theta(\theta_4-\frac{|\vecc{q}^{\bot}_2|}{x_2\varepsilon})\;\;
\Theta(\frac{|\vecc{q}^{\bot}_2|}{x_2\varepsilon}-\theta_2).
\end{eqnarray}
 In the case of a symmetrical angular acceptance (we restrict ourselves further
to this case only) we have:
\begin{eqnarray}
\theta_2=\theta_1,\quad \theta_4=\theta_3,\quad
\rho=\frac{\theta_3}{\theta_1} > 1.
\end{eqnarray}

We will present $\Sigma $ as the sum of various contributions:
\begin{eqnarray} \label{eq20}
\Sigma&=&\Sigma_0+\Sigma^{\gamma}+\Sigma^{2\gamma}
+\Sigma^{e^+e^-}+\Sigma^{3\gamma}+\Sigma^{e^+e^-\gamma}\\ \nonumber
&=&\Sigma_{00}(1+\delta_{0}+\delta^{\gamma}+\delta^{2\gamma}+
\delta^{e^+e^-}+\delta^{3\gamma}+\delta^{e^+e^-\gamma}),\;\\ \nonumber
\Sigma_{00}&=&1-\rho^{-2},
\end{eqnarray}
where $ \Sigma_0$ stands for a modified Born contribution, $\Sigma^{\gamma}$
for a contribution of one-photon emission (real and virtual) and so on.
The values of the $\delta^i$ as function of $x_c$ are given in Table~1
(see in Section 9). Being
stimulated by the representation in Eq.~(16), we shall
slightly modify the perturbation theory, using the full  propagator for the
$t$-channel
photon, which takes into account the growth of the electric charge at small
distances.
By integrating Eq.~(\ref{eq6}) with this convention, we obtain:
\begin{equation}
\Sigma_0=\theta_1^2\int\limits_{\theta_1^2}^{\theta_2^2}
\frac{\dd\theta^2}{\theta^4}
(1-\Pi(t))^{-2}+\Sigma_W+\Sigma_\theta,
\end{equation}
where $\Sigma_W$ is the correction due to the weak interaction:
\begin{equation}
\Sigma_W=\theta_1^2\int\limits_{\theta_1^2}^{\theta_2^2}\frac{\dd\theta^2}
{\theta^4}\delta_{\mbox{\tiny weak}}\, ,
\end{equation}
and the term $\Sigma_{\theta}$ comes from the expansion of the Born
cross--section in powers of $\theta^2$,
\begin{equation}
\Sigma_{\theta}=\theta_1^2\int\limits_{1}^{\rho^2}\frac{\dd z}{z}
(1-\Pi(-zQ_1^2))^{-2}\biggl(-\frac{1}{2}+z\theta_1^2\frac{9}{40}\biggr).
\end{equation}

The remaining contributions to $\Sigma$ in (\ref{eq20}) are considered below.

\section{Single hard-photon emission }

In order to calculate the contribution to $\Sigma$ due to the hard-photon
emission we start from  the corresponding differential cross--section
written in terms of energy fractions $x_{1,2}$ and transverse components
$\vecc{q}^{\bot}_{1,2}$ of the final particle momenta \cite{r9}:
\begin{eqnarray} \label{eq24}
\frac{\dd\sigma_{B}^{e^+e^-\rightarrow e^+e^-\gamma}}
{\dd x_1\dd^2\vecc{q}^{\bot}_1\dd x_2\dd^2\vecc{q}^{\bot}_2}&=&
\frac{2\alpha^3}{\pi^2} \;\biggl\{\frac{R(x_1;\vecc{q}^{\bot}_1,
\vecc{q^{\bot}}_2)
\;\delta(1-x_2)}{(\vecc{q}^{\bot}_2)^4\;(1-\Pi(-(\vecc{q}^{\bot}_2)^2))^2}
\\ \nonumber
&+& \frac{R(x_2;\vecc{q}^{\bot}_2,\vecc{q}^{\bot}_1)\;
\delta(1-x_1)}{(\vecc{q}^{\bot}_1)^4\;
(1-\Pi(-(\vecc{q}^{\bot}_1)^2))^2}\biggr\}\;(1+{\cal{O}}(\theta^2)),
\end{eqnarray}
where
\begin{eqnarray} \label{eq25}
&& R(x;\vecc{q}^{\bot}_1,\vecc{q}^{\bot}_2)=\frac{1+x^2}{1-x}\;
\biggl[{\frac{(\vecc{q}^{\bot}_2)^2(1-x)^2}{d_1d_2}-
\frac{2m^2(1-x)^2x}{1+x^2}\;\frac{(d_1-d_2)^2}{d_1^2d_2^2}}\biggr],
\\ \nonumber &&
d_1=m^2(1-x)^2+(\vecc{q}^{\bot}_1-\vecc{q}^{\bot}_2)^2,\qquad d_2=m^2(1-x)^2
+(\vecc{q}^{\bot}_1-x\vecc{q}^{\bot}_2)^2,
\end{eqnarray}
and we use the full photon propagator for the $t$-channel
photon.
Performing a simple azimuthal angle integration of Eq.~(24) we obtain for
the hard-photon emission the contribution $\Sigma^H$:
\begin{eqnarray} \label{eq26}
&& \Sigma^H=\frac{\alpha}{\pi}\int\limits_{x_c}^{1-\Delta}\dd x\;
\frac{1+x^2}{1-x}\; F(x,D_1,D_3;D_2,D_4 ),
\\ \nonumber &&
\end{eqnarray}
with
\begin{eqnarray} \label{eq27}
F=\int\limits_{D_1}^{D_3}\dd z_1\int\limits_{D_2}^{D_4}\frac{\dd z_2}{z_2}
(1-\Pi(-z_2Q_1^2))^{-2}
\biggl\{ \frac{1-x}{z_1-xz_2}(a_1^{-\frac{1}{2}}-xa_2^{-\frac{1}{2}})
- \frac{4x\sigma^2}{1+x^2}
\bigl[a_1^{-\frac{3}{2}}+x^2a_2^{-\frac{3}{2}} \bigr] \biggr\},
\end{eqnarray}
where
\begin{eqnarray}\label{eq28}
a_1=(z_1-z_2)^2+4z_2\sigma^2, \;\;\;a_2=(z_1-x^2z_2)^2+4x^2z_2\sigma^2,\;\;\;
\sigma^2=\frac{m^2}{Q_1^2}(1-x)^2,
\end{eqnarray}
and the integration limits in (27) in the symmetrical case are:
\begin{eqnarray} \label{eq29}
D_1=x^2,\quad D_2=1,\quad D_3=x^2\rho^2,\quad D_4=\rho^2.
\end{eqnarray}
From Eqs.~(\ref{eq26})--(\ref{eq29}) we have that:
\begin{eqnarray} \label{eq30}
\Sigma^H &=& \frac{\alpha}{\pi} \int\limits_{x_c}^{1-\Delta}\dd x
\frac{1+x^2}{1-x} \int\limits_{1}^{\rho^2}
\frac{\dd z}{z^2}(1-\Pi(-zQ_1^2))^{-2} \nonumber \\
&\times& \biggl\{ [1 +\Theta(x^2\rho^2-z)]\;
(L-1)+k(x,z) \biggr\},
\\ \nonumber
k(x,z)&=& \frac{(1-x)^2}{1+x^2} \;[1+\Theta(x^2\rho^2-z)]
+ \; L_1+\Theta(x^2\rho^2-z)\;L_2
%\\ \nonumber
+\Theta(z-x^2\rho^2) L_3 \, ,
\end{eqnarray}
where $L=\ln(zQ_1^2/m^2)$ and
\begin{eqnarray} \label{l123}
L_1&=&\ln\left|\frac{x^2(z-1)(\rho^2-z)}{(x-z)(x\rho^2-z)}\right|, \qquad
L_2=\ln\left|\frac{(z-x^2)(x^2\rho^2-z)}{x^2(x-z)(x\rho^2-z)}\right|,
\\ \nonumber
L_3&=&\ln\left|\frac{(z-x^2)(x\rho^2-z)}{(x-z)(x^2\rho^2-z)}\right|.
\end{eqnarray}
It is seen from Eq.~(\ref{eq30}) that $\Sigma^H$ contains  the  auxiliary parameter
$\Delta$. This parameter disappears, as it should, in the sum
$\Sigma^{\gamma} = \Sigma^H+\Sigma^{V+S}$, where $\Sigma^{V+S}$ is the
contribution of virtual and soft real photons which can be obtained using
Eq.~(15):
\begin{eqnarray} \label{eq32}
\Sigma^{\gamma}&=&\frac{\alpha}{\pi}\int\limits_{1}^{\rho^2}
\frac{\dd z}{z^2}\int\limits_{x_c}^{1}
\dd x (1-\Pi(-zQ_1^2))^{-2}\;\biggl\{
(L-1) P(x)
\\ \nonumber &\times&
[1+\Theta(x^2\rho^2-z)]+\frac {1+x^2}{1-x}k(x,z)-
\delta(1-x) \biggr\},
\end{eqnarray}
where
\begin{eqnarray} \label{eq33}
P(x)=\biggl(\frac{1+x^2}{1-x}\biggr)_+=\lim_{ \Delta \to 0 }\;
\biggl\{ \frac{1+x^2}{1-x}\;
\theta(1-x-\Delta)+(\frac{3}{2}+2\ln\Delta)\;\delta(1-x) \biggr\}
\end{eqnarray}
is the non--singlet splitting kernel (see Appendix A for details).

\section{Radiative corrections to ${\cal{O}}(\alpha^2)$}

A systematic treatment of all ${\cal O}(\alpha^2)$ contributions
is absent up to now. This is mainly due to the extreme complexity of
the analysis (more then 100 Feynman diagrams are to be taken into
account considering elastic and inelastic processes).
Nevertheless in the case of small scattering angles we may restrict
ourselves by considering only diagrams of the scattering type.
It is enough to make some rough estimates of other
contributions. Contributions of pure annihilation--type diagrams,
describing some ${\cal O}(\alpha^2)$ RC,
have so-called double--logarithmical enhancement~\cite{gorsh}  but,
fortunately, it is proportional to the 4th power of the small
scattering angle:
\begin{eqnarray}
(\Sigma^{\gamma\gamma})_{annih} \sim (\Sigma^{e^+e^-})_{annih}
\sim \theta^4(\frac{\alpha}{\pi})^2{\cal L}^4.
\end{eqnarray}
The contribution of interference of the scattering--type
and the annihilation--type amplitudes can be estimated as
\begin{eqnarray}
(\Sigma^{\gamma\gamma})_{interf} \sim (\Sigma^{e^+e^-})_{interf}
\sim \theta^2(\frac{\alpha}{\pi})^2\ln^4(\frac{Q^2}{s}).
\end{eqnarray}
The absence of large lograrithms here has a similar origin as in
(15). The uncertainties comes from the discussed contributions
are considered in sect.~9.

We consider first virtual two--loop corrections $d\sigma^{(2)}_{VV}$ to the
elastic scattering cross--section. Using the representation (\ref{eq16})
and the loop expansion
for the Dirac form factor $F_1$
\begin{eqnarray} \label{eq34}
F_1 = 1+\frac{\alpha}{\pi}F^{(1)}_1+\bigl(\frac{\alpha}{\pi}\bigr)^2F^{(2)}_1
\end{eqnarray}
one obtains
\begin{eqnarray} \label{eq35}
\frac{\dd\sigma^{(2)}_{VV}}{\dd c}=\frac{\dd\sigma_0}{\dd c}
\bigl(\frac{\alpha}{\pi}\bigr)^2
(1-\Pi(t))^{-2}\bigl[\, 6(F^{(1)}_1)^2 + 4F^{(2)}_1 \bigr].
\end{eqnarray}
The one-loop contribution to the form factor is well known:
\begin{eqnarray} \label{eq36}
F^{(1)}_1=(L-1)\ln\frac{\lambda}{m}+\frac{3}{4}L
- \frac{1}{4}L^2-1+\frac{1}{2}\zeta_2.
\end{eqnarray}
The two--loop correction can be obtained from the results of Ref.~\cite{r10}.
Let us present it
in the form
\begin{eqnarray}
F^{(2)}_1 = F^{\gamma\gamma}_1 + F^{e^+e^-}_1,
\end{eqnarray}
where the contribution $F^{e^+e^-}_1$ is related to the  vacuum
polarization by $e^+e^-$ pairs:
\begin{eqnarray} \label{eq38}
F^{e^+e^-}_1&=&-\frac{1}{36}L^3 +\frac{19}{72}L^2-
\biggl(\frac{265}{216}+\frac{1}{6}\zeta_2\biggr)L+{\cal O}(1), \\ \label{eq39}
F^{\gamma\gamma}_1&=&\frac{1}{32}L^4-\frac{3}{16}L^3
+ \biggl(\frac{17}{32}
- \frac{1}{8}\zeta_2\biggr)L^2 + \biggl(-\frac{21}{32}
- \frac{3}{8}\zeta_2+\frac{3}{2}\zeta_3\biggr)L
\\ \nonumber
&+& \frac{1}{2}(L-1)^2\ln^2 \frac{m}{\lambda}
+ (L-1)\biggl[-\frac{1}{4}L^2
+ \frac{3}{4}L-1+\frac{1}{2}\zeta_2\biggr]\ln\frac{\lambda}{m}
+{\cal O}(1), \\ \nonumber
\zeta_2&=&\sum\limits_{1}^{\infty}\frac{1}{n^2}=\frac{\pi^2}{6}\, ,\qquad
\zeta_3=\sum\limits_{1}^{\infty}\frac{1}{n^3}\approx 1.202 \, .
\end{eqnarray}
The photon mass $\lambda$ entering Eqs.~(\ref{eq36})--(\ref{eq39})
is cancelled in the
expression $\dd\sigma^{(2)}/\dd c\ $ for the sum of the virtual
and soft-photon corrections of the second order
$\dd\sigma^{(2)}_{VV}/\dd c$ (see Eq.~(\ref{eq35})),
$\dd\sigma^{(2)}_{SS}/\dd c\ $ and $\dd\sigma^{(2)}_{SV}/\dd c$.

The cross--section
$\dd\sigma^{(2)}_{SS}/\dd c\ $ for the emission of two soft photons,
each of energy smaller than $\Delta \varepsilon=\varepsilon \Delta$, is
$(\Delta \ll 1)$:
\begin{eqnarray} \label{eq40}
\dd\sigma^{(2)}_{SS} &=& \dd\sigma_0 \;\bigl(\frac{\alpha}{\pi}\bigr)^2
(1-\Pi(t))^{-2}
\;8 \biggl[ (L-1)\ln\frac{m\Delta}{\lambda}
+ \frac{1}{4}L^2 - \frac{1}{2} \zeta_2\biggr]^2,
\end{eqnarray}
and the virtual correction $\dd \sigma^{(2)}_{SV}/\dd c$ to the
cross--section of the single soft-photon emission is:
\begin{eqnarray} \label{eq41}
\dd\sigma^{(2)}_{SV} &=& \dd\sigma_0 \;\bigl(\frac{\alpha}{\pi}\bigr)^2
(1-\Pi(t))^{-2}
16F^{(1)}_1\biggl[(L-1)\ln \frac{m\Delta}{\lambda}
+\frac{1}{4}L^2-\frac{1}{2}\zeta_2\biggr].
\end{eqnarray}
The contribution to $\Sigma$ of this sum, except the part
coming
from $F^{e^+e^-}_1$ connected with the vacuum polarization,
contains no more than a second power  of $L$.
It  has the following form:
\begin{eqnarray} \label{eq42}
\Sigma_{S+V}^{\gamma\gamma}=\Sigma_{VV}
+\Sigma_{VS}+\Sigma_{SS}=\bigl(\frac{\alpha}{\pi}\bigr)^2
\int\limits_{1}^{\rho^2}\frac{\dd z}{z^2}
(1-\Pi(-zQ_1^2))^{-2}R_{S+V}^{\gamma\gamma}\, .
\end{eqnarray}
It is convenient to separate the $R_{S+V}^{\gamma\gamma}$ in the following
way:
\begin{eqnarray} \label{eq43}
R_{S+V}^{\gamma\gamma}&=&r_{S+V}^{\gamma\gamma}+r_{S+V\gamma\gamma}+
r_{S+V\gamma}^{\gamma}\, ,
\\ \nonumber
r_{S+V}^{\gamma\gamma}&=& r_{S+V\gamma\gamma}
= L^2 \biggl(2 \ln^2 \Delta + 3 \ln \Delta + \frac{9}{8}\biggr) \\ \nonumber
&+& L \biggl(- 4 \ln^2 \Delta - 7 \ln \Delta +3 \zeta_3- \frac{3}{2} \zeta_2
- \frac{45}{16}\biggr), \\ \nonumber
r_{S+V\gamma}^{\gamma}&=&4\bigl[(L-1)\ln\Delta+{3\over 4}L-1\bigr]^2.
\end{eqnarray}

The contribution to $\Sigma$ coming from $F^{e^+e^-}_1$ contains an $L^3$ term,
which is also cancelled when we take into account the soft pair
production contribution
\begin{eqnarray} \label{eq44}
\dd\sigma_S^{e^+e^-}&=&\bigl(\frac{\alpha}{\pi}\bigr)^2 \dd\sigma_0\;
(1-\Pi(t))^{-2} R_S^{e^+e^-}
\; =\; \bigl(\frac{\alpha}{\pi}\bigr)^2 \dd\sigma_0\; (1-\Pi(t))^{-2}
\biggl[ \frac{1}{9} (L+2\ln\Delta)^3 \\ \nonumber
&-& \frac{5}{9}(L+2\ln\Delta)^2+\biggl(\frac{56}{27}-\frac{2}{3}\zeta_2\biggr)
(L+2\ln\Delta)+{\cal O}(1)\biggl].
\end{eqnarray}
Thus for the contribution of the virtual and soft e$^+$ e$^-$ pairs to $\Sigma$
we have
\begin{eqnarray} \label{eq45}
\Sigma_{S+V}^{e^+e^-}&=&\bigl(\frac{\alpha}{\pi}\bigr)^2
\int\limits_{1}^{\rho^2}\,
\frac{\dd z}{z^2} (1-\Pi(-zQ_1^2))^{-2}R_{S+V}^{e^+e^-}\, , \\
R_{S+V}^{e^+e^-}&=&R_S^{e^+e^-}+4F_1^{e^+e^-}=L^2\biggl(\frac{2}{3}\ln\Delta
+\frac{1}{2}\biggr) + L\biggl(-\frac{17}{6}+\frac{4}{3}\ln^2\Delta \nonumber
\\ \nonumber
&-& \frac{20}{9}\ln\Delta-\frac{4}{3}\zeta_2\biggr)+{\cal O}(1).
\end{eqnarray}
In expressions (\ref{eq43})--(\ref{eq45}), $\Delta= \delta \varepsilon /
\varepsilon $ is the energy fraction carried by the soft pair, and it is assumed that
$2m \ll\delta\varepsilon\ll\varepsilon$.
Here we have taken into account only e$^+$ e$^-$ pair production.
The order of magnitude of the radiative correction due to pair production is
less than  $0.5\%$. A  rough estimate of the muon  pair contribution gives
less than  $0.05\%$ since $\ln(Q^2/m^2)\sim 3\ln(Q^2/m_{\mu}^2)$.
Contributions of pion and tau lepton pairs give corrections that are
still smaller.
Therefore, within the $0.1\%$  accuracy, we may omit any pair production
contribution except the e$^+$ e$^-$ one.

\section{Virtual and soft corrections
%\\
to the hard-photon emission}

By evaluating the corrections arising from the emission of virtual and real
soft photons which accompany a single hard-photon we will consider two cases.
The first case corresponds to the emission of the photons by the same fermion.
The second one occurs when the hard-photon is emitted by another fermion:

\begin{eqnarray} \label{eq46}
\dd\sigma\bigg|_{H(S+V)}=\dd\sigma^{H(S+V)} + \dd\sigma_{H(S+V)}
+ \dd\sigma^{H}_{(S+V)} + \dd\sigma_{H}^{(S+V)}.
\end{eqnarray}
In the case when both fermions emit, one finds that:
\begin{eqnarray} \label{eq47}
\Sigma^{H}_{(S+V)}+\Sigma_{H}^{(S+V)}=2\Sigma^{H}
\bigl(\frac{\alpha}{\pi}\bigr)\biggl[(L-1)\ln\Delta + \frac{3}{4}L-1\biggr],
\end{eqnarray}
where $\Sigma^H$ is given in Eq.~(\ref{eq30}).
A more complex expression arises when the radiative corrections
are applied to the same fermion line.
Here the cross--section may be expressed in terms
of the Compton tensor with an off--shell photon~\cite{r11}, which
describes the process
\begin{eqnarray} \label{eq48}
\gamma^{*}(q) + e^-(p_1)\rightarrow e^-(q_1) + \gamma(k)
+ (\gamma_{\mbox{\tiny soft}}).
\end{eqnarray}

In the limit of small--angle photon emission we have:
\begin{eqnarray} \label{eq48a}
\dd\sigma^{H(S+V)}&=& \frac{\alpha^4\dd x\dd^2\vecc{q}^{\bot}_1
\dd^2\vecc{q}^{\bot}_2}
{4x(1-x)(\vecc{q}^{\bot}_2)^4\pi^3}
\bigl[\bigl(B_{11}(s_1,t_1)+x^2B_{11}(t_1,s_1)\bigr)h+T\bigr], \\ \nonumber
T &=& T_{11} (s_1,t_1) + x^2 T_{11}(t_1,s_1)+x\bigl(T_{12}(s_1,t_1)
+ T_{12}(t_1,s_1)\bigr),
\\ \nonumber
h &=& 2 \biggl(L-\ln \frac{(\vecc{q}^{\bot}_2)^2}{-u_1}-1\biggr)
(2\ln \Delta-\ln x)+3L -\ln^2 x - \frac{9}{2}\, ,
\end{eqnarray}
where $\Delta=(\Delta \varepsilon/\varepsilon)\ll 1$,
$\Delta \varepsilon $ is the maximal energy of the soft photon,
escaping the detectors,
$B_{11}(s_1,t_1)=(-4(\vecc{q}^{\bot}_2)^2)/(s_1t_1)-8m^2/s_1^2$
is the Born Compton tensor component, and the invariants
are: $s_1=2q_1k,\ $  $t_1=-2p_1k,\ $  $u_1=(p_1-q_1)^2,\ $
$s_1+t_1+u_1=q^2$.

The final result (see Appendix~C for details) has the form:
\begin{eqnarray} \label{eq49}
\Sigma^{H(S+V)}&=&\Sigma_{H(S+V)}= \frac{1}{2}({\alpha \over \pi})^2
\int\limits_1^{\rho^2}\frac{\dd z}{z^2}
\int\limits_{x_c}^{1-\Delta}\frac{\dd x(1+x^2)}{1-x}\;L
\;\biggl\{\biggl(2\ln\Delta - \ln x
+ \frac{3}{2}\biggr) \\ \nonumber
&\times& [(L-1)(1+\Theta)+k(x,z)]
+\frac{1}{2}\ln^2x+(1+\Theta)[-2+\ln x-2\ln\Delta]
\\ \nonumber
&+& (1-\Theta)\biggl[\frac{1}{2} L \ln x+2\ln\Delta \ln x-\ln x\ln(1-x)
\\ \nonumber
&-& \ln^2x - \mbox{Li}_2(1-x) - \frac{x(1-x)+4x\ln x}{2(1+x^2)}\biggr]
- \frac{(1-x)^2}{2(1+x^2)}\biggr\}, \\ \nonumber
\mbox{Li}_2(x)&\equiv&-\int\limits_{0}^{x}\frac{\dd t}{t}\ln(1-t),
\end{eqnarray}
where $k(x,z)$ is given in Eq.~(\ref{eq30}) and
$\Theta \equiv \Theta (x^2\rho^2-z)$.

\section{Double hard-photon bremsstrahlung}

We now consider the contribution given by the process of emission of two hard
photons. We will distinguish two cases:  a) the double simultaneous
bremsstrahlung in opposite directions along electron and positron momenta,
and b) the double bremsstrahlung in the same direction along electron or
positron momentum.
The differential cross--section in the first case can be obtained by
using the factorization property of cross--sections within the impact parameter
representation \cite{r12}. It takes the following form \cite{r9}
(see Appendix A):
\begin{eqnarray} \label{eq50}
\frac{\dd\sigma^{e^+e^-\rightarrow (e^+\gamma)(e^-\gamma)}}
{\dd x_1\dd^2\vecc{q}^{\bot}_1\dd x_2\dd^2\vecc{q}^{\bot}_2}
=\frac{\alpha^4}{\pi^3}\int\limits_{}^{}
\frac{\dd^2\vecc{k}^{\bot}}{\pi (\vecc{k}^{\bot})^4}\;
(1-\Pi(-(\vecc{k}^{\bot})^2))^{-2}
R(x_1;\vecc{q}^{\bot}_1,\vecc{k}^{\bot})
R(x_2;\vecc{q}^{\bot}_2,-\vecc{k}^{\bot}),
\end{eqnarray}
where $R(x;\vecc{q}^{\bot},\vecc{k}^{\bot})$ is given by Eq.~(\ref{eq25}).
The calculation of the corresponding contribution $ \Sigma^H_H $ to $ \Sigma $
is analogous to the case of the single
hard-photon emission and the result has the form:
\begin{eqnarray} \label{eq52}
\Sigma_H^H=\frac{1}{4}\bigl(\frac{\alpha}{\pi}\bigr)^2\int\limits_{0}^{\infty}
\frac{\dd z}{z^2}(1-\Pi(-zQ_1^2))^{-2}\int\limits_{x_c}^{1-\Delta}\dd x_1
\int\limits_{x_c/x_1}^{1-\Delta}\dd x_2\;
\frac{1+x_1^2}{1-x_1}\;\frac{1+x_2^2}{1-x_2}\;\Phi(x_1,z)\Phi(x_2,z),
\end{eqnarray}
where (see Eq.~(\ref{l123})):
\begin{eqnarray} \label{eq53}
\Phi(x,z)&=&(L-1)[\Theta(z-1)\Theta(\rho^2-z)+\Theta(z-x^2)
\Theta(\rho^2x^2-z)]
\\ \nonumber
&+&L_3[-\Theta(x^2-z)+\Theta(z-x^2\rho^2)]
+\biggl(L_2+\frac{(1-x)^2}{1+x^2}\biggr)
\Theta(z-x^2)\Theta(x^2\rho^2-z) \\ \nonumber
&+& \biggl(L_1+\frac{(1-x)^2}{1+x^2}\biggr)
\Theta(z-1)\Theta(\rho^2-z) \\ \nonumber
&+& (\Theta(1-z) -\Theta(z-\rho^2))
\ln\left|\frac{(z-x)(\rho^2-z)}{(x\rho^2-z)(z-1)}\right|.
\end{eqnarray}

Let us now turn to  the double hard-photon emission in the
same direction and the hard e$^+$ e$^-$ pair production.
Here we use the method  developed by one of us \cite{r13,r14}.
We will distinguish the
collinear and semi--collinear kinematics of final particles.
In the first case all produced particles move in the cones
within the polar angles $\theta_i<
\theta_0\ll 1$ centered along the charged-particle momenta (final or initial).
In the semi--collinear region only one produced particle moves inside
those cones,
while the other moves outside them. For the totally inclusive
cross--section, such a distinction no longer has physical meaning and the
dependence on the auxiliary parameter $\theta_0$ disappears.
We underline that in this way all double and single--logarithmical
contributions may be extracted rigorously. The contribution of the
region when both the photons move outside the small cones does not
contain any large logarithm $L$. The systematic omission of those
contributions in the double bremsstrahlung and pair production processes
is the source of uncertainties of order
$(\alpha/\pi)^2\leq 0.6\cdot 10^{-5}$.

The contribution of both collinear and semi--collinear regions
(we consider for definiteness
the emission of both hard photons along the electron, since the
contribution of the emission along the positron is the same) has
the form (see Appendix~B for details):
\begin{eqnarray} \label{eq54}
\Sigma^{HH}&=&\Sigma_{HH}=\frac{1}{4}\bigl(\frac{\alpha}{\pi}\bigr)^2
\int\limits_{1}^{\rho^2}\frac{\dd z}{z^2}(1-\Pi(-zQ_1^2))^{-2} \\ \nonumber
&\times& \int\limits_{x_c}^{1-2\Delta}\dd x
\int\limits_{\Delta}^{1-x-\Delta}\dd x_1
\frac{I^{HH}L}{x_1(1-x-x_1)(1-x_1)^2}, \\ \nonumber
I^{HH}&=&A\;\Theta(x^2\rho^2-z)+B+C\;\Theta((1-x_1)^2\rho^2-z),
\end{eqnarray}
where
\begin{eqnarray}
A&=&\gamma\beta\biggl(\frac{L}{2}+\ln \frac{(\rho^2 x^2-z)^2}
{x^2(\rho^2 x (1-x_1)-z)^2}\biggr)+(x^2+(1-x_1)^4)\ln\frac{(1-x_1)^2(1-x-x_1)}{xx_1}
+\gamma_A,
\nonumber \\
B&=&\gamma\beta\left(\frac{L}{2}+\ln\left|\frac{x^2(z-1)(\rho^2-z)(z-x^2)
(z-(1-x_1)^2)^2(\rho^2x(1-x_1)-z)^2}
{(\rho^2 x^2-z)(z-(1-x_1))^2(\rho^2(1-x_1)^2-z)^2(z-x(1-x_1))^2}\right|\right)
\nonumber \\ \label{eq55}
&+&(x^2+(1-x_1)^4)\ln\frac{(1-x_1)^2x_1}{x(1-x-x_1)}+\delta_B,
\nonumber \\  \nonumber
C&=&\gamma\beta\left(L+2\ln\left|\frac{x(\rho^2 (1-x_1)^2-z)^2}{(1-x_1)^2
(\rho^2 x (1-x_1)-z)(\rho^2(1-x_1)-z)}\right|\right) \\ \nonumber
&-& 2(1-x_1)\beta - 2x(1-x_1)\gamma,
\end{eqnarray}
where
\begin{eqnarray*}
\gamma&=&1+(1-x_1)^2,\qquad \beta=x^2+(1-x_1)^2, \\
\gamma_A&=&xx_1(1-x-x_1)-x_1^2(1-x-x_1)^2 -2(1-x_1)\beta , \nonumber \\
\delta_B&=&xx_1(1-x-x_1)-x_1^2(1-x-x_1)^2 -2x(1-x_1)\gamma. \nonumber
\end{eqnarray*}

One may see that the combinations
\begin{eqnarray} \label{eq56}
r^{\gamma\gamma}+\Sigma^{H(S+V)}+\Sigma^{HH}, \qquad
r^{\gamma}_{\gamma}+\Sigma^{H}_{S+V}+\Sigma_H^{S+V}+\Sigma^{H}_{H}
\end{eqnarray}

with  $r^{\gamma\gamma}$ and  $r^{\gamma}_{\gamma}$
normalized (see Eqs.~(42,43)) to

\begin{eqnarray} \label{eq56a}
r^{\gamma\gamma} \rightarrow
(\frac{\alpha}{\pi})^2\int\limits_{1}^{\rho^2}
\frac{\dd z}{z^2}(1-\Pi(-zQ_1^2))^{-2} r_{S+V}^{\gamma\gamma} \nonumber ,
\end{eqnarray}
and
\begin{eqnarray} \label{eq56b}
r^{\gamma}_{\gamma} \rightarrow
(\frac{\alpha}{\pi})^2\int\limits_{1}^{\rho^2}
\frac{\dd z}{z^2}(1-\Pi(-zQ_1^2))^{-2} r^{\gamma}_{S+V\gamma} \nonumber,
\end{eqnarray}
respectively, do not depend on $\Delta$ for $\Delta\rightarrow0$
(see Appendix~E).

The total expression $ \Sigma^{2\gamma} $, which describes the
contribution to (\ref{eq20}) from the two--photon (real and virtual)
 emission processes is determined by expressions (43), (47), (49), (51)
, (53) and (55). Furthermore it does not depend on the auxiliary parameter
$\Delta$ and has the form:
\begin{eqnarray} \label{eq57}
\Sigma^{2\gamma}&=&\Sigma_{S+V}^{\gamma\gamma}+2\Sigma^{H(V+S)}
+ 2\Sigma^{H}_{S+V} + \Sigma_H^H+2\Sigma^{HH} \\ \nonumber
&=&\Sigma^{\gamma\gamma}+\;\Sigma_{\gamma}^{\gamma}+
(\frac{\alpha}{\pi})^2{\cal{L}}(\phi^{\gamma\gamma}
+\phi^{\gamma}_{\gamma}), \qquad
{\cal{L}}=\ln\frac{\varepsilon^2\theta_1^2}{m^2}.
\end{eqnarray}
The leading contributions $ \Sigma^{\gamma\gamma},
\Sigma_{\gamma}^{\gamma} $ have the following forms (see Appendix~D):
\begin{eqnarray} \label{eq58}
\Sigma^{\gamma\gamma}&=&\frac{1}{2}\bigl(\frac{\alpha}{\pi}\bigr)^2
\int\limits_{1}^{\rho^2}\frac{\dd z}{z^2}L^2
(1-\Pi(-Q_1^2z))^{-2}\int\limits_{x_c}^{1}\dd x\;\biggl\{
\frac{1}{2}P^{(2)}(x)\;[\;\Theta(x^2\rho^2-z)+1] \nonumber \\
&+& \int\limits_{x}^{1}\frac{\dd t}{t}
P(t)\;P(\frac{x}{t})\;\Theta(t^2\rho^2-z)\biggr\}, \\  \label{eq59}
P^{(2)}(x)&=&\int\limits_{x}^{1}\frac{\dd t}{t}P(t)\;P(\frac{x}{t})=
\lim_{\Delta \to 0} \biggl\{\;\biggl[\biggl(2\ln\Delta+\frac{3}{2}\biggr)^2
- 4\zeta_2\biggr]\;\delta(1-x) \\ \nonumber
&+& 2\biggl[\frac{1+x^2}{1-x}\biggl(2\ln(1-x)-\ln x + \frac{3}{2}\biggr)
+\frac{1}{2}(1+x)\ln x- 1+x\biggr]\;\Theta(1-x-\Delta)\biggr\},
 \nonumber \\ \label{eq60}
\Sigma_{\gamma}^{\gamma}&=&\frac{1}{4}\bigl(\frac{\alpha}{\pi}\bigr)^2
\int\limits_{0}^{\infty}\frac{\dd z}{z^2}L^2
(1-\Pi(-Q_1^2z))^{-2}\int\limits_{x_c}^{1}\dd x_1
\int\limits_{x_c/x_1}^{1}\dd x_2 P(x_1)P(x_2) \\ \nonumber
&\times& \bigl[\Theta(z-1)\Theta(\rho^2-z)
+ \Theta(z-x_1^2)\Theta(x_1^2\rho^2-z)\bigr] \\ \nonumber
&\times& \bigl[\Theta(z-1)\Theta(\rho^2-z)
+ \Theta(z-x_2^2)\Theta(x_2^2\rho^2-z)\bigr].
\end{eqnarray}

We see that the leading contributions to $\Sigma^{2\gamma}$ may be expressed
in terms of kernels for the evolution equation for structure functions.

The functions $\phi^{\gamma\gamma}$ and $\phi^{\gamma}_{\gamma}$ 
in expression Eq.~(\ref{eq57}) collect the next-to-leading
contributions which cannot be obtained by the
structure functions method \cite{r15}. They have a form that can be obtained
by comparing the results in the leading logarithmic approximation
with the logarithmic ones given above.

\section{Pair production}

Pair production process in high--energy e$^+$ e$^-$ collisions
was considered
about 60 years ago (see \cite{r9} and references therein).
In particular it was found that the total cross--section
contains cubic terms in large logarithm $L$. These terms come
from the kinematics when the scattered electron and positron
move in narrow (with opening angles $\sim m/\epsilon$) cones
and the created pair have the invariant mass of the order
of $m$ and moves preferably along either the electron
beam direction or the positron one. According to the conditions
of the LEP detectors, such a kinematics can be excluded.
In the relevant kinematical region a parton-like description
could be used giving $L^2$ and $L$-enhanced terms.

We accept the LEP~1 conventions whereby an event of the Bhabha process
is defined as one in which
the angles of the simultaneously registered particles hitting
opposite detectors (see Eq.~(\ref{deli0})).

The method, developed by one of us (N.P.M.) \cite{r13,r14},
of calculating the real hard pair production cross--section
within logarithmic accuracy (see the discussion in sect.~6)
consists in separating the
contributions of the collinear and semi--collinear kinematical regions.
In the first one (CK)  we suggest that both electron and positron from
the created pair go in the narrow cone around the direction of one
charged particle [the projectile (scattered) electron $\vecc{p}_1$
$(\vecc{q}_1)$ or the projectile (scattered) positron $\vecc{p}_2$
$(\vecc{q}_2)]$:
\begin{eqnarray} \label{p1}
\widehat{\vecc{p}_+\vecc{p}_-} \sim \widehat{\vecc{p}_-\vecc{p}_i}
\sim \widehat{\vecc{p}_+\vecc{p}_i}
< \theta_0 \ll 1, \qquad \varepsilon\theta_0/m \gg 1, \qquad
\vecc{p}_i=\vecc{p}_1,\,\vecc{p}_2,\,\vecc{q}_1,\,\vecc{q}_2\, .
\end{eqnarray}
The contribution of the CK contains terms of order
$(\alpha L/\pi)^2$, $(\alpha/\pi)^2L\ln(\theta_0/\theta)$
and $(\alpha/ \pi)^2L$, where $\theta=\widehat{\vecc{p}_-\vecc{q}_1}$ is
the scattering angle.
In the semi--collinear region
only one of conditions (\ref{p1}) on the angles is fulfilled:
\begin{eqnarray} \label{p2}
&& \widehat{\vecc{p}_+\vecc{p}_-} < \theta_0,\quad
\widehat{\vecc{p}_{\pm}\vecc{p}_i} > \theta_0\, ; \quad \mbox{or} \quad
\widehat{\vecc{p}_-\vecc{p}_i} < \theta_0,\quad
\widehat{\vecc{p}_+\vecc{p}_i} > \theta_0\, ; \\ \nonumber
&& \mbox{or} \quad
\widehat{\vecc{p}_-\vecc{p}_i} > \theta_0,\quad
\widehat{\vecc{p}_+\vecc{p}_i} < \theta_0\, .
\end{eqnarray}
The contribution of the SCK contains terms of the form:
\begin{eqnarray} \label{p3}
\left( \frac{\alpha}{\pi} \right) ^2 L \ln\frac{\theta_0}{\theta},
\qquad \left( \frac{\alpha}{\pi} \right) ^2 L.
\end{eqnarray}
The auxiliary parameter $\theta_0$ drops out in the total sum of
the CK and SCK contributions.

All possible mechanisms for pair creation (singlet and
non--singlet) and the identity of the particles in
the final state are taken into account \cite{r18}. In the case of small--angle
Bhabha scattering only a part of the total 36 tree-type Feynman
diagrams are relevant, i.e. the scattering diagrams\footnote{We have verified
that the interference between the amplitudes describing the production of
pairs moving in the electron direction and the positron one cancels.
This is known as up--down (interference) cancellation \cite{r18}.}.

The sum of the contributions due to virtual pair emission
(due to the vacuum polarization insertions in the virtual photon
Green's function) and of those due to the real soft pair emission
does not contain cubic $(\sim L^3)$ terms but depends on the auxiliary
parameter $\Delta = \delta\varepsilon/\varepsilon$
$(m_e\ll\delta\varepsilon\ll\varepsilon$, where
$\delta\varepsilon$ is the sum of the energies of the soft pair components).
The $\Delta$-dependence disappears in the total sum after
the contributions due to real hard pair production are added. Before summing
one has to integrate the hard pair contributions over the energy
fractions of the pair components, as well as over those of the scattered
electron and positron:
\begin{eqnarray} \label{p4}
\Delta&=&\frac{\Delta\varepsilon}{\varepsilon}< x_1+x_2, \qquad
x_c<x=1-x_1-x_2<1-\Delta, \\ \nonumber
x_1&=&\frac{\varepsilon_+}{\varepsilon}, \qquad
x_2=\frac{\varepsilon_-}{\varepsilon}, \qquad
x  =\frac{q_1^0}{\varepsilon}\, ,
\end{eqnarray}
where $\varepsilon_{\pm}$ are the energies of the positron and electron from
the created pair. We consider for definiteness the case when the created hard
pair moves close to the direction of the initial (or scattered) electron.

Consider first the collinear kinematics.
There are four different CK regions, when the created pair goes
in the direction of the incident (scattered) electron or positron.
We will consider only two of them, corresponding to the initial and
the final electron directions. For the case of pair emission
parallel to the initial electron, it is useful to decompose the particle
momenta into longitudinal and transverse components:
\begin{eqnarray} \label{p5}
p_+&=&x_1p_1 + p_+^{\bot}, \qquad p_-=x_2p_1 + p_-^{\bot},
\qquad q_1=xp_1 + q_1^{\bot}, \\ \nonumber
x&=&1-x_1-x_2, \qquad q_2 \approx p_2, \qquad
p_+^{\bot} + p_-^{\bot} + q_1^{\bot}=0,
\end{eqnarray}
where $p_i^{\bot}$ are the two--dimensional momenta of the final
particles, which are transverse with respect to the initial
electron beam direction.
It is convenient to introduce dimensionless
quantities for the relevant kinematical invariants:
\begin{eqnarray} \label{p6}
z_i&=&\left(\frac{\varepsilon\theta_i}{m}\right)^2, \qquad
0<z_i< \left(\frac{\varepsilon\theta_0}{m}\right)^2 \gg 1, \\ \nonumber
A&=&\frac{(p_++p_-)^2}{m^2}=(x_1x_2)^{-1}\bigl[
(1-x)^2+x_1^2x_2^2(z_1+z_2-2\sqrt{z_1z_2}\cos\phi )\bigr], \\ \nonumber
A_1&=&\frac{2p_1p_-}{m^2}=x_2^{-1}\bigl[1+x_2^2+x_2^2z_2\bigr], \qquad
A_2=\frac{2p_1p_+}{m^2}=x_1^{-1}\bigl[1+x_1^2+x_1^2z_1\bigr], \\ \nonumber
C&=&\frac{(p_1-p_-)^2}{m^2}=2-A_1,
\qquad D=\frac{(p_1-q_1)^2}{m^2}-1=A-A_1-A_2,
\end{eqnarray}
where $\phi$ is the azimuthal angle between the
$(\vecc{p}_1 \vecc{p}_+^{\bot})$ and $(\vecc{p}_1 \vecc{p}_-^{\bot})$ planes.

Keeping only the terms from the sum over spin states
of the square of the absolute value of the matrix element,
which give non--zero contributions
to the cross--section in the limit $\theta_0 \to 0$, we
find that only 8 from the total 36 Feynman diagrams
are essential \cite{r18}.

The result has the factorized form:
\begin{eqnarray} \label{p7}
\sum\limits_{\mbox{\tiny spins}} |M|^2 \Big|_{p_+,p_-\parallel p_1}
=\sum\limits_{\mbox{\tiny spins}} |M_0|^2 \, 2^7 \pi^2 \alpha^2 \frac{I}{m^4}\, ,
\end{eqnarray}
where one of the multipliers corresponds to the matrix element
in the Born approximation (without pair production):
\begin{eqnarray} \label{p8}
&& \sum\limits_{\mbox{\tiny spins}} |M_0|^2=2^7\pi^2\alpha^2
\left(\frac{s^4+t^4+u^4}{s^2t^2}\right), \\ \nonumber
&& \qquad s=2p_1p_2, \qquad t=-Q^2x,
\qquad u=-s-t,
\end{eqnarray}
and the quantity $I$, which stands for the collinear factor,
coincides with the expression for $I_a$ obtained in \cite{r14}.
We write it here in terms of our kinematical variables:
\begin{eqnarray} \label{p9}
I&=&(1-x_2)^{-2} \left(\frac{A(1-x_2)+Dx_2}{DC}\right)^2
+ (1-x)^{-2} \left(\frac{C(1-x)-Dx_2}{AD}\right)^2 \\ \nonumber
&+&\frac{1}{2xAD}\left[ \frac{2(1-x_2)^2-(1-x)^2}{1-x}
+ \frac{x_1x-x_2}{1-x_2} + 3(x_2-x) \right] \\ \nonumber
&+& \frac{1}{2xCD}\biggl[ \frac{(1-x_2)^2-2(1-x)^2}{1-x_2}
+ \frac{x-x_1x_2}{1-x} + 3(x_2-x) \biggr] \\ \nonumber
&+& \frac{x_2(x^2+x_2^2)}{2x(1-x_2)(1-x)AC} + \frac{3x}{D^2}
+ \frac{2C}{AD^2} + \frac{2A}{CD^2} + \frac{2(1-x_2)}{xA^2D} \\ \nonumber
&-& \frac{4C}{xA^2D^2} - \frac{4A}{D^2C^2}
+ \frac{1}{DC^2}\left[ \frac{(x_1-x)(1+x_2)}{x(1-x_2)}
- 2\frac{1-x}{x}\right].
\end{eqnarray}
We rewrite the phase volume of the final particles as
\begin{eqnarray} \label{p10}
\dd \Gamma&=&\frac{\dd^3q_1\dd^3q_2}{(2\pi)^6 2q_1^02q_2^0}(2\pi)^4
\delta^{(4)}(p_1x+p_2-q_1-q_2)\\ \nonumber  &\times&
m^42^{-8}\pi^{-4}x_1x_2\dd x_1\dd x_2
\dd z_1 \dd z_2 \frac{\dd \phi}{2\pi}\, .
\end{eqnarray}
Using the table of integrals given in Appendix~F
we further integrate over the variables of the created pair.
Following a similar procedure in the case when the pair moves
in the direction of the scattered electron, integrating the resulting
sum over the energy fractions of the pair components, and finally adding
the contribution of the two remaining CK regions (when the pair goes
in the positron directions), we obtain\footnote{
Some misprints, which occur in the expressions for $f(x)$ and
$f_1(x)$ in \cite{r14,r18}, are corrected here.}:

\begin{eqnarray} \label{p12}
\dd \sigma_{\mbox{\tiny coll}} &=& \frac{\alpha^4\dd x}{\pi Q_1^2}
\int\limits_{1}^{\rho^2}\!\!\!\;\;\frac{\dd z}{z^2}\;L\;
\biggl\{R_0(x)\left(L+2\ln\frac{\lambda^2}{z}\right)
(1+\Theta) \\ \nonumber
&+& 4R_0(x)\ln x + 2\Theta f(x) + 2f_1(x) \biggr\}\, , \qquad
\lambda=\frac{\theta_0}{\theta_{\mbox{\tiny min}}},\\ \nonumber
\Theta &\equiv& \Theta(x^2\rho^2-z)=
\left\{ \begin{array}{l}  1,\quad x^2\rho^2 > z, \\ 0,\quad x^2\rho^2 \leq z,
\end{array} \right. \\ \nonumber
R_0(x)&=&\frac{2}{3}\,\frac{1+x^2}{1-x} + \frac{(1-x)}{3x}
(4+7x+4x^2) + 2(1+x)\ln x, \\ \nonumber
f(x)&=& - \frac{107}{9} + \frac{136}{9}x - \frac{2}{3}x^2 - \frac{4}{3x}
- \frac{20}{9(1-x)} + \frac{2}{3} \bigl[ - 4x^2 - 5x + 1 \\ \nonumber
&+& \frac{4}{x(1-x)} \bigr] \ln (1-x) + \frac{1}{3} \bigl[
8x^2+5x-7-\frac{13}{1-x} \bigr] \ln x
- \frac{2}{1-x}\ln^2x \\ \nonumber
&+& 4(1+x)\ln x \ln(1-x) - \frac{2(3x^2-1)}{1-x}\mbox{Li}_2(1-x), \\ \nonumber
f_1(x)&=&-x\;\mbox{Re}\; f(\frac{1}{x})= - \frac{116}{9}
+ \frac{127}{9}x + \frac{4}{3}x^2
+ \frac{2}{3x} - \frac{20}{9(1-x)} + \frac{2}{3} \bigl[ - 4x^2
\\ \nonumber
&-& 5x + 1 + \frac{4}{x(1-x)} \bigr] \ln (1-x)
+ \frac{1}{3} \bigl[8x^2-10x-10+\frac{5}{1-x} \bigr] \ln x
\\ \nonumber
&-& (1+x)\ln^2x + 4(1+x)\ln x \ln(1-x)
- \frac{2(x^2-3)}{1-x}\mbox{Li}_2(1-x), \\ \nonumber
\qquad Q_1&=&\varepsilon\theta_{\mbox{\tiny min}},
\qquad L=\ln\frac{zQ^2_1}{m^2}\, .
\end{eqnarray}

Consider now semi--collinear kinematical regions.
We will restrict ourselves again to the case in which the created pair
goes close to the electron momentum (initial or final).
A similar treatment applies in the CM system in the
case in which the pair follows the positron momentum.
There are three different semi--collinear regions, which
contribute to the cross--section within the required
accuracy. The first region includes the events for which
the created pair has very small invariant mass:
\begin{eqnarray*}
4m^2 \ll (p_++p_-)^2 \ll |q^2|,
\end{eqnarray*}
and the pair escapes the narrow cones (defined by $\theta_0$)
in both the incident and scattered electron momentum directions.
We will refer to this SCK region as $\vecc{p}_+\parallel\vecc{p}_-$.
The reason is the smallness (in comparison with $s$) of the square of the
four--momentum of the virtual photon converting to the pair \cite{r18}.

The second SCK region includes the events for which the invariant mass
of the created positron and the scattered electron is small,
$4m^2 \ll (p_++q_1)^2 \ll |q^2|$, with the restriction that the positron
should escape the narrow cone in the initial electron momentum
direction. We refer to it as $\vecc{p}_+\parallel\vecc{q}_1$ \cite{r18}.

The third SCK region includes the events in which the created electron
goes inside the narrow cone in the initial electron momentum
direction, but the created positron does not. We refer to it as
$\vecc{p}_-\parallel\vecc{p}_1$ \cite{r18}.

The differential cross--section takes the following form:
\begin{eqnarray} \label{eq:16}
\dd \sigma &=& \frac{\alpha^4}{8\pi^4 s^2}
\, \frac{|M|^2}{q^4} \, \frac{\dd x_1 \dd x_2 \dd x}{x_1x_2x}
\dd^2\vecc{p}_+^{\bot} \dd^2\vecc{p}_-^{\bot} \dd^2\vecc{q}_1^{\bot}
\dd^2\vecc{q}_2^{\bot}
\delta(1-x_1-x_2-x) \\ \nonumber &\times &
\delta^{(2)}(\vecc{p}_+^{\bot} + \vecc{p}_-^{\bot} +\vecc{q}_1^{\bot} +
\vecc{q}_2^{\bot})\, ,
\end{eqnarray}
where $x_1$ $(x_2)$, $x$ and $\vecc{p}_+^{\bot}$ $(\vecc{p}_-^{\bot})$,
$\vecc{q}_1^{\bot}$
are the energy fractions and the perpendicular momenta of the created
positron (electron) and the scattered electron (positron) respectively;
$s=(p_1+p_2)^2$ and $q^2=-Q^2=(p_2-q_2)^2=-\varepsilon^2\theta^2$
are the center-of-mass energy squared and the momentum transferred
squared; the matrix element squared $|M|^2$  takes different forms
according to the different SCK regions.

For the differential cross--section in the $\vecc{p}_+\parallel\vecc{p}_-$
region we have
(see, for details, \cite{r17}):
\begin{eqnarray} \label{eq:22}
\dd \sigma_{\vecc{p}_+\parallel\vecc{p}_-} &=& \frac{\alpha^4}{\pi} L
\;\dd x \;\dd x_2 \frac{\dd (\vecc{q}_2^{\bot})^2}{(\vecc{q}_2^{\bot})^2}
\frac{\dd (\vecc{q}_1^{\bot})^2}{(\vecc{q}_1^{\bot}+\vecc{q}_2^{\bot})^2}
\\ \nonumber &\times&
\frac{\dd \phi}{2\pi} \frac{1}{(\vecc{q}_1^{\bot}+x\vecc{q}_2^{\bot})^2}
\biggl[(1-x_1)^2+(1-x_2)^2-\frac{4xx_1x_2}{(1-x)^2} \biggr],
\end{eqnarray}
where $\phi$ is the angle between the two--dimensional vectors
$\vecc{q}_1^{\bot}$ and $\vecc{q}_2^{\bot}$, $\vecc{q}_{1,2}^{\bot}$
are the transverse
momentum components of the final electrons, $x_{1,2}$ are their
energy fractions ($x=1-x_1-x_2$).
At this stage it is necessary to use the restrictions on the
two--dimensional momenta $\vecc{q}_1^{\bot}$ and $\vecc{q}_2^{\bot}$.
They appear when the
contribution of the CK region (which here represents the narrow
cones with opening angle $\theta_0 $ in the momentum directions
of both incident and scattered electrons) is excluded:
\begin{equation} \label{eq:23}
\left|\frac{\vecc{p}_+^{\bot}}{\varepsilon_+}\right| > \theta_0, \qquad
\left|\vecc{r}^{\bot}\right|=\left|\frac{\vecc{p}_+^{\bot}}{\varepsilon_+}
-\frac{\vecc{q}_1^{\bot}}{\varepsilon_2}\right| > \theta_0 \, ,
\end{equation}
where $\varepsilon_+ $ and $ \varepsilon_2 $ are the energies of the
created positron and the scattered electron respectively.
In order to exclude $p_+^{\bot}$ from the above equation
we use the conservation of the perpendicular momentum, in this
case:
\begin{eqnarray*}
\vecc{q}_1^{\bot}+\vecc{q}_2^{\bot}+\frac{1-x}{x_1}\vecc{p}_+^{\bot}=0.
\end{eqnarray*}

In the semi--collinear region $\vecc{p}_+\parallel\vecc{q}_1$
we obtain:
\begin{eqnarray} \label{eq:30}
\dd \sigma_{\vecc{p}_+\parallel\vecc{q}_1} &=& \frac{\alpha^4}{\pi} L\;
\dd x\; \dd x_2\; \frac{\dd (\vecc{q}_2^{\bot})^2}{(\vecc{q}_2^{\bot})^2}
\frac{\dd (\vecc{q}_1^{\bot})^2}{(\vecc{q}_1^{\bot})^2}
\\ \nonumber &\times&
\frac{\dd \phi}{2\pi} \frac{1}{(\vecc{q}_1^{\bot}+x\vecc{q}_2^{\bot})^2}
\frac{x^2}{(1-x_2)^2}
\biggl[ (1-x)^2+(1-x_1)^2-\frac{4xx_1x_2}{(1-x_2)^2} \biggr],
\end{eqnarray}
with the restrictions
\begin{eqnarray} \label{eq:31}
\left| \frac{\vecc{p}_-^{\bot}}{\varepsilon_-}
- \frac{\vecc{q}_1^{\bot}}{\varepsilon_2} \right| > \theta_0, \qquad
\vecc{p}_-^{\bot} + \vecc{q}_2^{\bot} + \frac{1-x_2}{x}\vecc{q}_1^{\bot}=0.
\end{eqnarray}

Finally for the $\vecc{p}_-\parallel\vecc{p}_1$
semi--collinear region we get:
\begin{eqnarray} \label{eq:35}
\dd \sigma_{\vecc{p}_-\parallel\vecc{p}_1} &=& \frac{\alpha^4}{\pi} L\;
\dd x\; \dd x_2\; \frac{\dd (\vecc{q}_2^{\bot})^2}{(\vecc{q}_2^{\bot})^2}
\frac{\dd (\vecc{q}_1^{\bot})^2}{(\vecc{q}_1^{\bot})^2}
\\ \nonumber &\times&
\frac{\dd \phi}{2\pi} \frac{1}{(\vecc{q}_1^{\bot}+\vecc{q}_2^{\bot})^2}
\biggl[\frac{(1-x)^2+(1-x_1)^2}{(1-x_2)^2}-\frac{4xx_1x_2}{(1-x_2)^4} \biggr].
\end{eqnarray}

The restriction due to the exclusion
of the collinear region when the created pair moves inside a
narrow cone in the direction of the initial electron has the form
\begin{eqnarray} \label{eq:36}
\frac{|\vecc{p}_+^{\bot}|}{\varepsilon_1} > \theta_0, \qquad
\vecc{p}_+^{\bot} + \vecc{q}_1^{\bot} + \vecc{q}_2^{\bot}=0.
\end{eqnarray}

In order to obtain the finite expression for the cross--section
we have to add $\dd \sigma_{\vecc{p}_+\parallel\vecc{p}_-}
+ \dd \sigma_{\vecc{p}_+\parallel\vecc{q}_1}
+ \dd \sigma_{\vecc{p}_-\parallel\vecc{p}_1}$
to the contribution of the
collinear kinematics region (\ref{p12})
and those due to the production
of virtual and soft pairs. Taking into account the leading and
next-to-leading terms we can write the full hard pair contribution
including also the pair emission along the positron direction,
after the integration over $x_2$ as
\begin{eqnarray} \label{eq:41}
\sigma_{\mbox{\tiny hard}}&=&2\;\frac{\alpha^4}{\pi Q_1^2}\int\limits_{1}^{\rho^2}
\frac{\dd z}{z^2} \int\limits_{x_c}^{1-\Delta}\!\dd x
\biggl\{ L^2(1+\Theta)R(x) + {\cal L}[\Theta F_1(x)+F_2(x)] \biggr\},
\\ \nonumber
F_1(x)&=&d(x)+C_1(x),\qquad F_2(x)=d(x)+C_2(x),
\\ \nonumber
d(x)&=&\frac{1}{1-x}\biggl(\frac{8}{3}\ln(1-x)-\frac{20}{9}\biggr),
\\ \nonumber
C_1(x)&=& - \frac{113}{9} + \frac{142}{9}x - \frac{2}{3}x^2
- \frac{4}{3x} - \frac{4}{3}(1+x)\ln(1-x)
\\ \nonumber
&+& \frac{2}{3}\frac{1+x^2}{1-x} \biggl[
\ln\frac{(x^2\rho^2-z)^2}{(x\rho^2-z)^2} - 3\mbox{Li}_2(1-x) \biggr]
+ \bigl(8x^2+3x-9-\frac{8}{x}
\\ \nonumber
&-&\frac{7}{1-x} \bigr) \ln x + \frac{2(5x^2-6)}{1-x}\ln^2x
+ \beta(x)\ln\frac{(x^2\rho^2-z)^2}{\rho^4},
\\ \nonumber
C_2(x)&=& - \frac{122}{9} + \frac{133}{9}x + \frac{4}{3}x^2
+ \frac{2}{3x} - \frac{4}{3}(1+x)\ln(1-x)
\\ \nonumber
&+& \frac{2}{3}\frac{1+x^2}{1-x} \biggl[
\ln \left| \frac{(z-x^2)(\rho^2-z)(z-1)}{(x^2\rho^2-z)(z-x)^2} \right|
+ 3\mbox{Li}_2(1-x) \biggr]
\\ \nonumber
&+& \frac{1}{3}\bigl(-8x^2-32x-20+\frac{13}{1-x}+\frac{8}{x} \bigr) \ln x
+ 3(1+x)\ln^2x
\\ \nonumber
&+& \beta(x)\ln\left|\frac{(z-x^2)(\rho^2-z)(z-1)}{x^2\rho^2-z}\right|,
\qquad \beta = 2R(x) - \frac{2}{3}\frac{1+x^2}{1-x}\, ,\\
R(x)&=&\frac{1}{3} \frac{1+x^2}{1-x} + \frac{1-x}{6x}(4+7x+4x^2)
+ (1+x)\ln x.
\end{eqnarray}

Eq.~(\ref{eq:41}) describes the small--angle
high--energy cross--section
for the pair production process, provided that the created hard pair 
can move along both electron and positron beam directions. 

The contribution to the cross--section of the small--angle Bhabha scattering
connected with the real soft (with energy lower than
$\Delta  \varepsilon$) and virtual pair production can be defined
\cite{r18} by the formula:
\begin{eqnarray} \label{eq:45}
\sigma_{\mbox{\tiny soft+virt}}&=&\frac{4\alpha^4}{\pi Q_1^2}
\int\limits_{1}^{\rho^2}\;\;\frac{\dd z}{z^2} \biggl\{L^2
\biggl(\frac{2}{3}\ln\Delta + \frac{1}{2}\biggr)
+ {\cal L}\biggl(-\frac{17}{6}+\frac{4}{3}\ln^2\Delta
\\ \nonumber
&-& \frac{20}{9} \ln \Delta - \frac{4}{3} \zeta_2\biggr) \biggr\}.
\end{eqnarray}
Using Eqs.~(\ref{eq:41}) and (\ref{eq:45}) it is easy to verify
that the auxiliary parameter $\Delta$ is cancelled in the sum
$\sigma_{\mbox{\tiny pair}}=\sigma_{\mbox{\tiny hard}}
+\sigma_{\mbox{\tiny soft+virt}}$.
We can, therefore, write the total contribution $\sigma_{\mbox{\tiny pair}}$ as
\begin{eqnarray} \label{eq:46}
&& \sigma_{\mbox{\tiny pair}}=\frac{2\alpha^4}{\pi Q_1^2}
\int\limits_{1}^{\rho^2}\;\;
\frac{\dd z}{z^2} \biggl\{ L^2\bigl(1+\frac{4}{3}\ln(1-x_c)
-\frac{2}{3}\int\limits_{x_c}^{1}\frac{\dd x}{1-x}\bar{\Theta}\bigr)
+ {\cal L}\biggl[-\frac{17}{3}
\\ \nonumber && \quad
- \frac{8}{3}\zeta_2
-\frac{40}{9}\ln(1-x_c)+\frac{8}{3}\ln^2(1-x_c)
+ \int\limits_{x_c}^{1}\frac{\dd x}{1-x}\bar{\Theta}\cdot
\bigl(\frac{20}{9}-\frac{8}{3}\ln(1-x)\bigr) \biggr]
\\ \nonumber && \quad
+ \int\limits_{x_c}^{1}\;\!\dd x\bigl[ L^2(1+\Theta)\bar{R}(x)
+ {\cal L}(\Theta C_1(x) + C_2(x)) \bigr] \biggr\},
\quad \bar{R}(x)=R(x)-\frac{2}{3(1-x)},
\\ \nonumber && \quad
\bar{\Theta}=1-\Theta.
\end{eqnarray}

The right-hand side of Eq.~(\ref{eq:46}) gives the contribution to
the small--angle Bhabha scattering cross--section for
pair production. It is finite and can be used for numerical
estimations. The leading term can be described by the
electron structure function $D_e^{\bar{e}}(x)$ \cite{r16}.

\section{Terms of ${\cal O}(\alpha{\cal{L}})^3 $ }

In order to evaluate the leading logarithmic contribution
represented by terms of the type $(\alpha {\cal {L}})^3$,
we use the iteration up to $\beta^3 $ of the
master equation~\cite{r15} obtained in Ref.~\cite{r16}.
To simplify the analytical
expressions we adopt here a realistic assumption about the smallness
of the threshold for the detection of the hard subprocess energy
and neglect terms of the order of:
\begin{eqnarray} \label{q1}
x_c^n(\frac{\alpha}{\pi}{\cal L})^3 \leq 3\cdot 10^{-5},\qquad
n=1,2,3\, .
\end{eqnarray}
We may, therefore, limit ourselves to consider the emission by the
initial electron and positron. Three photons (virtual and real)
contribution to $\Sigma$ have the form:
\begin{eqnarray} \label{q2}
\Sigma^{3\gamma}&=&\frac{1}{4}\;(\frac{\alpha}{\pi}{\cal{L}})^3
\int\limits_{1}^{\rho^2}\frac{\dd z}{z^2}
\int\limits_{x_c}^{1}\dd x_1
\int\limits_{x_c}^{1}\dd x_2\;\Theta(x_1x_2-x_c)\;
\biggl[\frac{1}{6}\delta(1-x_2)\;P^{(3)}(x_1)
\\ \nonumber &\times&
\Theta(x_1^2\rho^2-z)
+\frac{1}{2x_1^2}P^{(2)}(x_1)P(x_2)\Theta_1\Theta_2\biggr]\;(1+{\cal O}(x_c^3)),
\end{eqnarray}
where $P(x)$  and  $P^{(2)}(x)$ are given by Eqs.~(\ref{eq33}) and (\ref{eq58})
correspondingly:
\begin{eqnarray} \label{q3}
\Theta_1\Theta_2&=&\Theta\bigl(z-\frac{x_2^2}{x_1^2}\bigr)\;
\Theta\bigl(\rho^2 \frac{x_2^2}{x_1^2}-z\bigr),
\nonumber \\
P^{(3)}(x)&=&\delta(1-x)\;\Delta_t+\Theta(1-x-\Delta)\;\Theta_t,
\nonumber\\
\Delta_t&=&48 \biggl[\frac{1}{3}\zeta_3 -\frac{1}{2}\zeta_2
\biggl(\ln \Delta + \frac{3}{4}\biggr)
+\frac{1}{6}\biggl(\ln \Delta + \frac{3}{4}\biggr)^3 \biggr],
\\ \nonumber
\Theta_t&=& 48 \biggl\{\frac{1}{2}\;\frac{1+x^2}{1-x}
\biggl[\frac{9}{32}-\frac{1}{2}\zeta_2
+\frac{3}{4}\ln(1-x)-\frac{3}{8}\ln x+\frac{1}{2}\ln^2(1-x)
\\ \nonumber
&+& \frac{1}{12} \ln^2x-\frac{1}{2}\ln x\ln (1-x)\biggr]
+\frac{1}{8}(1+x)\ln x\ln(1-x)-\frac{1}{4}
(1-x)\ln(1-x)
\\ \nonumber
&+& \frac{1}{32}(5-3x)\ln x -\frac{1}{16}(1-x)-\frac{1}{32}(1+x)\ln^2x
+\frac{1}{8}(1+x)\mbox{Li}_2(1-x)\biggr\}.
\end{eqnarray}
The contribution to $\Sigma$ of the process of pair production
accompanied by photon
emission when both, pair and photons, may be real and virtual has the
form (with respect to paper by M.~Skrzypek~\cite{r16}
we include also the non--singlet mechanism
of pair production):
\begin{eqnarray}
&& \Sigma^{e^+e^-\gamma}=\frac{1}{4}(\frac{\alpha}{\pi}
{\cal{L}})^3\int\limits_{1}^{\rho^2}\,
\dd z\;z^{-2}\int\limits_{x_c}^{1}\,\dd x_1\int\limits_{x_c}^{1}\,\dd x_2
\;\Theta(x_1x_2-x_c)\;
\nonumber \\ \nonumber && \qquad \times
\{\frac{1}{3}[R^P(x_1)-\frac{1}{3}R^s(x_1)]\;\delta(1-x_2)
\Theta(x_1^2\rho^2-z)+\frac{1}{2x_1^2}\;P(x_2)R(x_1)
\;\Theta_1\Theta_2\},
\end{eqnarray}
where
\begin{eqnarray}
&& R(x)=R^s(x)+\frac{2}{3}P(x), \qquad
R^s(x)=\frac{1-x}{3x}(4+7x+4x^2)+2(1+x)\ln x,
\\ \nonumber &&
R^P(x)=R^s(x)(\frac{3}{2}+2\ln(1-x))+(1+x)(-\ln^2x
+ 4\mbox{Li}_2(1-x) \\ \nonumber && \qquad
+\frac{1}{3}(-9-3x+8x^2)\ln x+
\frac{2}{3}(-\frac{3}{x}-8+8x+3x^2) + \frac{2}{3}P^{(2)}(x).
\end{eqnarray}
The total expression for $\Sigma$ in Eq.~(20) is the sum of
the contributions in Eqs.
(21), (32), (56), (60), (66) and (68).
The quantity $ \Sigma $ depends on the parameters $x_c,\rho$
and $Q_1^2$.

\section{Estimates of neglected terms and numerical results}

The uncertainty of our calculations is defined by neglected terms.
Let us list them.

a) Terms of the first order RC coming from annihilation--type
diagrams (15):
\begin{eqnarray}
\frac{\alpha}{\pi}\theta_1^2\int\limits_{\theta_1^2}^{\theta_2^2}
\frac{\dd \theta}{\theta^2}\;\Delta_\theta \leq 0.10\cdot 10^{-4}.
\end{eqnarray}

b) Similar terms in the second order do not exceed
(see sect.~4)
\begin{eqnarray}
&& (\frac{\alpha}{\pi})^2\theta_1^2
\int\limits_{\theta_1^2}^{\theta_2^2}\frac{\dd \theta}{\theta^2}\; l^4
\leq 0.23\cdot 10^{-4},\\ \nonumber
&& (\frac{\alpha}{\pi})^2(\theta_2^4-\theta_1^4){\cal L}^4\leq
0.5\cdot 10^{-5}.
\end{eqnarray}

c) We neglect terms which violate the eikonal approximation:
\begin{eqnarray}
\frac{\alpha}{\pi}\;\frac{Q^2}{s}\leq 0.3\cdot 10^{-6}.
\end{eqnarray}

d) We omit term of the second order which are not enhanced by
large logarithms:
\begin{eqnarray}
(\frac{\alpha}{\pi})^2= 0.5\cdot 10^{-5}.
\end{eqnarray}

e) Creation of heavy pairs $(\mu\mu$, $\tau\tau$, $\pi\pi$, $\dots)$ gives
in sum at least one order of magnitude smaller than the corresponding
contribution due to light particle production~\cite{rele}:
\begin{eqnarray}
\Sigma_{\pi\pi}+\Sigma_{\mu\mu}+\Sigma_{\tau\tau} \leq 0.1\;\Sigma^{e^+e^-}
\leq 0.5\cdot 10^{-4}.
\end{eqnarray}

f) Higher--order corrections, including soft and collinear multi-photon
contributions, can be neglected since they only give contributions
of the type $(\alpha L/\pi)^4\leq 0.2\cdot 10^{-5}$ or less.

g) The terms in the third order associated with the emission off
the final particles\footnote{Usually, in a calorimetric experimental
set--up such terms do not contribute.}:
\begin{eqnarray}
x_c(\frac{\alpha {\cal L}}{\pi})^3 \leq 0.3\cdot 10^{-4}\ \ \
(\mbox{for}\ x_c=0.5).
\end{eqnarray}

Regarding all the uncertainties a)--g) and (82) as independent ones
we conclude the total theoretical uncertainty of our results to be
$\pm 0.006\%$. "

Let us  define $\Sigma_0^0$ to be equal to $\Sigma_0|_{\Pi=0}$
(see Eq.~(21)), which corresponds to the Born cross--section
obtained by switching off the vacuum polarization contribution
$\Pi(t)$. For the experimentally observable cross--section we obtain:

\begin{eqnarray}\label{deli0}
\sigma=\frac{4\pi\alpha^2}{Q_1^2}\Sigma_0^0\;(1+\delta_0+\delta^{\gamma}
+\delta^{2\gamma}+\delta^{e^+e^-}+\delta^{3\gamma}+\delta^{e^+e^-\gamma}),
\end{eqnarray}
where
\begin{eqnarray}
\Sigma_0^0=\Sigma_0|_{\Pi=0}=1-\rho^{-2}+\Sigma_W+\Sigma_{\theta}|_{\Pi=0}
\end{eqnarray}
and
\begin{eqnarray}
\delta_0=\frac{\Sigma_0-\Sigma_0^0}{\Sigma_0^0}\;;\;
\delta^{\gamma}=\frac{\Sigma^{\gamma}}{\Sigma_0^0}\;;\;
\delta^{2\gamma}=\frac{\Sigma^{2\gamma}}{\Sigma_0^0}\;;\;
\cdots\;\;\;.
\end{eqnarray}

The numerical results are presented below in Table~1.

\vspace{.3cm}
{\bf Table~1:}  The values of $\delta^i$ in per cent
for  $\sqrt{s}=91.161$ GeV, $\theta_1=1.61^{\circ}$,
$\theta_2=2.8^{\circ}$, $\sin^2\theta_W=0.2283$,
$\Gamma_Z=2.4857$ GeV.
\vspace{.3cm}

\begin{tabular}{|c|c|c|c|c|c|c|c|c|}
\hline
$x_c$ & $\delta_0 $ & $\delta^{\gamma} $
&$\delta^{2\gamma}_{\mbox{\tiny leading}}$
&$\delta^{2\gamma}_{\mbox{\tiny nonleading}}$
& $\delta^{e^+e^-} $ & $\delta^{e^+e^-\gamma} $
&$\delta^{3\gamma} $&$\sum \delta^i $ \\ \hline
0.1& 4.120& $-$8.918& 0.657&  0.162& $-$0.016& $-$0.017
& $-$0.019& $-$4.031$\pm$0.006 \\
0.2& 4.120& $-$9.226& 0.636&  0.156& $-$0.027& $-$0.011
& $-$0.016& $-$4.368$\pm$0.006 \\
0.3& 4.120& $-$9.626& 0.615&  0.148& $-$0.033& $-$0.008
& $-$0.013& $-$4.797$\pm$0.006 \\
0.4& 4.120&$-$10.147& 0.586&  0.139& $-$0.039& $-$0.005
& $-$0.010& $-$5.356$\pm$0.006 \\
0.5& 4.120&$-$10.850& 0.539&  0.129& $-$0.044& $-$0.003
& $-$0.006& $-$6.115$\pm$0.006 \\
0.6& 4.120&$-$11.866& 0.437&  0.132& $-$0.049& $-$0.002
& $-$0.001& $-$7.229$\pm$0.006 \\
0.7& 4.120&$-$13.770& 0.379&  0.130& $-$0.057& $-$0.001
&    0.005& $-$9.194$\pm$0.006 \\
0.8& 4.120&$-$17.423& 0.608&  0.089& $-$0.069&    0.001
&    0.013&$-$12.661$\pm$0.006 \\
0.9& 4.120&$-$25.269&1.952&$-$0.085& $-$0.085&    0.005
&    0.017&$-$19.379$\pm$0.006 \\
\hline
\end{tabular}

\vskip 20.0pt

Each of these contributions to $\sigma$ has a sign that can change
because of the interplay between real and virtual corrections.
The cross--section corresponding to  the Born diagrams for producing
a real particle is always positive, whereas
the sign of the radiative corrections depends on the order of
perturbation theory. For the virtual corrections at odd orders it is
negative, and at even orders it is positive.
When the aperture of the counters is small the compensation between real
and virtual corrections is not complete.
In the limiting case of small aperture $(\rho\to 1,\ x_c\to 1)$
the virtual contributions dominate.
\vskip 1.pt
The numerical results were obtained by using the NLLBHA fortran code 
\cite{nllbha}. 
\vskip 1.pt

The analytical and the numerical calculations for the cross--section
in the non symmetrical Narrow-Wide configuration are in progress and
will be presented elsewhere.
\vskip 10.0pt

\newpage
\leftline{\bf Acknowledgements}
\vskip 10.0pt
We are grateful for support to the Istituto Nazionale 
di Fisica Nucleare (INFN), to the International Association (INTAS) 
for the grant 93-1867 and 
to the Russian Foundation for Fundamental Investigations (RFFI)
for the grant 96-02-17512.
One of us (L.T.) would like to thank H.~Czyz, M.~Dallavalle,
B.~Pietrzyk and T.~Pullia for several useful discussions at various stages
of the work and the CERN theory group for the hospitality. 
Three of us (A.A., E.K. and N.M.) would like to thank the INFN
Laboratori Nazionali
di Frascati, the Dipartimento di Fisica dell'Universit\'a di Roma
"Tor Vergata" and the Dipartimento di Fisica
dell'Universit\'a di Parma for their hospitality
during the preparation of this work.
One of us (A.A.) is thankful to the Royal Swedish Academy of Sciences for
an ICFPM grant.

\newpage
%-----------------------------------------------------
%\section*{References and comments}

\newpage

\def\beq#1{\begin{equation}\label{#1}}
\def\beeq#1{\begin{eqnarray}\label{#1}}
\def\eeq{\end{equation}}
\def\eeeq{\end{eqnarray}}

\appendix
\newpage
%\chapter{}
\section*{Appendix A}
{\bf Infinite momentum frame kinematics}
\vskip 20.0pt
\setcounter{equation}{0}
\renewcommand{\theequation}{A.\arabic{equation}}

In this Appendix we will consider the kinematics we use
to obtain the electron--positron and photon distributions.
Due to the peculiar range of momenta and angles of the
reaction, it is particularly convenient to use the Sudakov
parametrization or infinite momentum frame kinematics.
For the reaction
\beq{kin}
e^+(p_2)+e^-(p_1) \rightarrow e^+(q_2)+e^-(q_1)+\gamma(k)
\eeq
let us introduce the Sudakov decomposition:
\beeq{sud}
q_1&=&\alpha_1 \tilde{p}_2+\beta_1 \tilde{p}_1+q_{1}^{\bot},\qquad
q_2=\alpha_2 \tilde{p}_2+\beta_2 \tilde{p}_1+q_{2}^{\bot}
\nonumber \\
k&=&\alpha\tilde{p}_2+\beta\tilde{p}_1+k^{\bot},
\eeeq
where $\tilde{p}_{1,2}$ are almost light-like four--vectors:
\begin{eqnarray}
q_{i}^{\bot}p_1=q_{i}^{\bot}p_2=0,\qquad
(q_i^{\bot})^2= - (\vecc{q}_i^{\bot})^2<0,
\end{eqnarray}
\beeq{p11p22}
\tilde{p}_1&=& p_1-\frac{m^2}{s}p_2\, ,\quad
\tilde{p}_2=p_2-\frac{m^2}{s}p_1, \nonumber \\
p_1^2&=&p_2^2=q_1^2=q_2^2=m^2,\quad k^2=0, \quad
\tilde{p}_1^2=\tilde{p}_2^2=\frac{m^6}{s^2}\, ,
\nonumber  \\
s&=&2p_1p_2=2\tilde{p}_1\tilde{p}_2=2\tilde{p}_1p_2=2\tilde{p}_2p_1\gg m^2,
\eeeq
where $\vecc{q}_{i}^{\bot}$ are Euclidean two--dimensional
vectors in the center-of-mass reference frame.

We consider the kinematical configuration when the photon is emitted
in the direction close to the initial electron.
We have the mass-shell conditions:
\beeq{d}
q_1^2&=&s\alpha_1\beta_1-(\vecc{q}_1^{\bot})^2 =m^2,\qquad
\alpha_1=\frac{(\vecc{q}_1^{\bot})^2+ m^2}{s\beta_1}\, , \\ \nonumber
(q'_2)^2&=&s\alpha_2 \beta_2-(\vecc{q}_2^{\bot})^2 =m^2, \qquad
\beta_2=\frac{(\vecc{q}_2^{\bot})^2+m^2}{s\alpha_2}\, , \\ \nonumber
k^2&=&s\alpha\beta-(\vecc{k}^{\bot})^2=0,\qquad
s\alpha=\frac{(\vecc{k}^{\bot})^2}{\beta}\, , \\ \nonumber
\alpha_2&=&1,\quad |\beta_2|\sim |\alpha_1|\sim |\alpha|\ll 1,
\quad \beta_1\sim\beta\sim 1.
\eeeq

The components along $\tilde{p}_1$ of the jets containing $e^-(q_1)$
and $\gamma(k)$ have a value of ${\cal O}(1)$.
The phase volume decomposition with
$\dd^4q_1=\frac{s}{2}\;\dd\alpha_1\dd\beta_1\dd^2\vecc{q}_1^{\bot}$ is:
\beeq{phase}
\dd\phi&=&\frac{\dd^3\vecc{q}_1\dd^3\vecc{q}_2\dd^3\vecc{k}}
{2q_1^0 2q_2^0 2\omega}\delta^{(4)}(p_1+p_2-q_1-q_2-k) \\ \nonumber
&=& \frac{1}{4s\beta\beta_1}\dd\beta\dd\beta_1\delta(1-\beta-\beta_1)
\dd^2\vecc{k}^{\bot}\dd^2\vecc{q}_1^{\bot}\dd^2\vecc{q}_2^{\bot}
\delta^{(2)}(\vecc{q}_1^{\bot}
+ \vecc{q}_2^{\bot} + \vecc{k}^{\bot} ).
\eeeq

The conservation law reads (we introduce a new four--momentum $q$ of the
exchanged photon):
\beeq{llll}
p_1 + q = q_1+k,\qquad p_2=q_2+q.
\eeeq

The inverse propagators are (here and further we use $\beta_1=x$):

\beeq{prop}
&& (p_1-k)^2-m^2=\frac{-1}{1-x}d_1\, , \qquad
(p_1+q)^2-m^2=\frac{1}{x(1-x)}d,\\ \nonumber
&& q^2=-(\vecc{q}_2^{\bot})^2, \quad
d=m^2(1-x)^2+(\vecc{q}_1^{\bot} + \vecc{q}^{\bot} x)^2, \quad
d_1=m^2(1-x)^2+(\vecc{q}_1^{\bot} + \vecc{q}^{\bot})^2.
\eeeq

The matrix element reads
\beeq{matrix}
M&=&\frac{g^{\mu \nu}}{q^2} \bar{v}(p_2) \gamma_{\mu} v(q_2)
\bar{u}(q_1)O_{\nu}u(p_1)
\nonumber \\
O^{\nu}&=&\gamma^{\nu}\frac{\hat{p}_1-\hat{k}+m}{(p_1-k)^2-m^2}\hat{e}
+\hat{e}\frac{\hat{p}_1+\hat{q}+m}{(p_1+q)^2-m^2}\gamma^{\nu}.
\eeeq

The following decomposition of the metric tensor $g_{\mu\nu}$ is used:
\beeq{dec}
g_{\mu\nu}=g_{\mu\nu}^{\perp}+\frac {p_1^{\mu}p_2^{\nu}+p_1^{\nu}p_2^{\mu}}
{p_1p_2}
\simeq \frac{2p_1^{\mu}p_2^{\nu}}{s}\biggl(1 + {\cal O}
(\frac{\vecc{q}^{\bot 2}}{s}) \biggr).
\eeeq

We use also the identity

\beq{orho}
p_2^{\nu}\bar{u}(q_1)O_{\nu}u(p_1)\equiv \bar{u}(q_1)\hat{v}_{\rho}u(p_1)e_{\rho}(k).
\eeq

The generalized vertex $v_{\rho}$ has the form \cite{r9}:
\beeq{V}
v_{\rho}= s \gamma_{\rho} x(1-x) \left(\frac{1}{d} -\frac{1}{d_1}\right)
- \frac{ \gamma_{\rho} \hat{k} \hat{p_2}}{d} x (1-x) -\frac
{ \hat{p_2} \hat{k} \gamma_{\rho} } {d_1} (1-x).
\eeeq

The evaluation of the spin sum of the squared matrix element gives

\beq{spin}
\sum_{\mbox{\tiny spin}}|\bar{v}(q_2)\hat{p_1}v(p_2)|^2
=\mbox{Tr}\;\hat{p_2}\hat{p_1}\hat{p_2}\hat{p_1}=2s^2,
\eeq

The squared matrix element for the single photon radiation is given
by

\beeq{r}
R&=&-\frac{1}{4s^2}\; \mbox{Tr}\; (\hat{p_1}+m)\hat{v}_{\mu}(\hat{p_1}+\hat{k}
-\hat{q}+m)\hat{v}_{\mu} \\ \nonumber
&=& x[-2xm^2 (d-d_1)^2
+(\vecc{q}_2^{\bot})^2 (1+x^2)dd_1]\frac{1}{d^2d_1^2}\, .
\eeeq

Finally we obtain that
\beq{dis}
\dd\sigma^{e^+e^-\rightarrow e^+(e^-\gamma)}=2 \alpha^3
\frac{\dd^2\vecc{q}_1^{\bot}\dd\vecc{q}_2^{\bot}\dd x(1-x)}
{\pi^2 ((\vecc{q}_2^{\bot})^2)^2(dd_1)^2}
[-2xm^2(d-d_1)^2 + (\vecc{q}_2^{\bot})^2(1+x^2)dd_1].
\eeq

In the same way we may obtain the cross--section for the process of
the double bremsstrahlung in the opposite directions:
\beeq{doubopp}
\frac{d\sigma^{e^+e^- \rightarrow e^+ \gamma e^- \gamma}}
{\dd^2\vecc{q}_1^{\bot}\dd^2\vecc{q}_2^{\bot}\dd x_1\dd x_2} &=&
\frac{\alpha^4(1+x_1^2)(1+x_2^2)}{\pi^4(1-x_1)(1-x_2)}
\int \frac{\dd^2\vecc{q}^{\bot}}{((\vecc{q}^{\bot})^2)^2} \\ \nonumber
&\times& \biggl[\frac{(\vecc{q}^{\bot})^2(1-x_1)^2}{d_1d_2}
- \frac{2 x_1}{1 + x_1^2}\frac{m^2(1-x_1)^2(d_1-d_2)^2}{d_1^2 d_2^2}\biggr]
\\ \nonumber &\times&
\biggl[ \frac{(\vecc{q}^{\bot})^2(1-x_2)^2} {\tilde{d}_1 \tilde{d}_2}
- \frac {2x_2}
{1+x_2^2} \frac {m^2(1-x_2)^2(\tilde{d}_2-\tilde{d}_1)^2}
{\tilde{d}_1^2 \tilde{d}_2^2 }\biggr],
\eeeq
where $x_1$, $\vecc{q}_1^{\bot}$ and $x_2$, $\vecc{q}_2^{\bot}$ are
the energy fractions and the components
transverse to the beam
axis of the scattered electron and positron, respectively;
$\vecc{q}^{\bot}$ is the transverse two--dimensional momentum of
the exchanged photon;
\beeq{deq}
d_1&=&(1-x_1)^2m^2+(\vecc{q}_1^{\bot} - \vecc{q}^{\bot}x_1)^2, \quad
d_2=(1-x_1)^2m^2+(\vecc{q}_1^{\bot} -  \vecc{q}^{\bot})^2, \\ \nonumber
\tilde{d}_1&=&(1-x_2)^2m^2+(\vecc{q}_2^{\bot} + \vecc{q}^{\bot}x_2)^2, \quad
\tilde{d}_2=(1-x_2)^2m^2+(\vecc{q}_2^{\bot} + \vecc{q}^{\bot})^2.
\eeeq

Let us now discuss the restrictions on the $\dd^2\vecc{q}_1^{\bot}$,
$\dd^2\vecc{q}_2^{\bot}$ integration imposed
by experimental conditions of the electron and positron tagging.
We consider the emission of a hard photon along the electron direction.
We will consider the symmetric case:
\begin{eqnarray}
&& \theta_1<\theta_e=\frac{|\vecc{q}_1^{\bot}|}{x\varepsilon}<\theta_2\, ,
\qquad \theta_e=\hat{\vecc{p}_1\vecc{q}_1}, \\ \nonumber
&& \theta_1<\theta_{\bar{e}}=\frac{|\vecc{q}_2^{\bot}|}{\varepsilon}
<\theta_2\, , \qquad
\theta_{\bar{e}}=\hat{\vecc{p}_2\vecc{q}_2}.
\end{eqnarray}

Here $\theta_1$ and $\theta_2$ are the minimal and maximal angles of
aperture for the counters. It is convenient to introduce dimensionless
quantities $\rho=\theta_2/\theta_1$,
$z_{1,2}=(\vecc{q}_{1,2})^2/Q_1^2$  $~(Q_1=\varepsilon \theta_1)$.

The region in the $z_1,z_2$ plane that gives the largest contribution
to $\Sigma$ is made by two narrow strips along the lines $z_1=z_2$
 and $z_1=x^2z_2$. Therefore the leading logarithmic contribution will appear
only in the cases where these lines cross the rectangle defined by
$x^2<z_1<\rho^2x^2\;,\;1<z_2<\rho^2$.
Note that the line $z_1=x^2z_2$, which corresponds to the emission of one hard
photon along the momentum of the scattered electron, is the diagonal of the
rectangle defined above. As for the line $z_1=z_2$, which corresponds to the
emission along the initial electron momentum, it crosses the rectangle only
if $x^2\rho^2>z_2\;,\;x\rho>1$. This last condition defines the appearance
of leading contributions to $\Sigma^H$.

For the contribution from the photon emission by the initial
electron we have:
\begin{eqnarray}
F_1&=&\Theta(1-\rho x)\int\limits_{1}^{\rho^2}\frac{\dd z_2}{z_2^2}
\int\limits_{x^2}^{x^2\rho^2}\frac{\dd z_1\; z_2(1-x)}{(z_1-xz_2)(z_2-z_1)}
\nonumber \\ \nonumber
&+& \Theta(x\rho-1)
\int\limits_{x^2\rho^2}^{\rho^2}\frac{\dd z_2}{z^2_2}
\int\limits_{x^2}^{x^2\rho^2}\frac{\dd z_1\; z_2(1-x)}{(z_1-xz_2)(z_2-z_1)}
\\ \nonumber
&+& \Theta(x\rho-1)\int\limits_{1}^{x^2\rho^2}\frac{\dd z_2}{z^2_2}
\biggl\{\int\limits_{x^2}^{z_2-\eta}\frac{\dd z_1\; z_2(1-x)}{(z_1-xz_2)(z_2-z_1)}
+ \int\limits_{z_2+\eta}^{x^2\rho^2}\frac{\dd z_1\; z_2(1-x)}{(z_1-xz_2)(z_1-z_2)}
\\
&+& \int\limits_{z_2-\eta}^{z_2+\eta}\frac{\dd z_1}{\sqrt{R}}
- \frac{2x\sigma^2}{1+x^2}\int\limits_{z_2-\eta}^{z_2+\eta}
\frac{2\dd z_1\; z_2}{\sqrt{R^3}} \biggr\},
\quad R=(z_2-z_1)^2 + 4\sigma^2z_2,
\end{eqnarray}
where we introduced the auxiliary parameter $\eta$, $~\sigma^2\ll\eta\ll 1$.
Neglecting the terms of order $\eta$ we obtain:
\begin{eqnarray}
F_1&=&\int\limits_{1}^{\rho^2}\frac{\dd z}{z^2}\biggl\{
\Theta(\rho x-1)\Theta(x^2\rho^2-z)\biggl(L-\frac{2x}{1+x^2}\biggr)
\\ \nonumber
&+& \Theta(x^2\rho^2-z)L_2 + \Theta(z-x^2\rho^2)L_3
\biggr\},
\end{eqnarray}
where $L_i$ are given in eq.~(\ref{l123}) and we used the identity
$\Theta(1-\rho x) +  \Theta(\rho x-1)\Theta(z-x^2\rho^2)= \Theta(z-x^2\rho^2)$.

In the same way we obtain for the final electron emission:
\begin{eqnarray}
F_2&=&\int\limits_{1}^{\rho^2}\frac{\dd z}{z^2}\biggl\{
L - \frac{2x}{1+x^2} + L_1 \biggr\}.
\end{eqnarray}

The total contribution due to one hard photon emission in small--angle
Bhabha scattering therefore reads:

\begin{eqnarray}
\Sigma^H=\frac{\alpha}{\pi}\int\limits_{x_c}^{1-\Delta}
\dd x\; \frac{1+x^2}{1-x}(F_1+F_2).
\end{eqnarray}

%\newpage
%\end{document}
\newpage
%\chapter{}
\section*{Appendix B}
{\bf The contribution to $\Sigma$ from the semi--collinear region
of emission \\ of two hard photons in the same direction}
\vskip 20.0pt
\setcounter{equation}{0}
\renewcommand{\theequation}{B.\arabic{equation}}

An alternative way to use the quasi-real electron approximation is
to compute the cross--section directly. To logarithmic accuracy
we may restrict ourselves to considering only two regions
i) the one with photon with momentum $k_1$ emitted along the
momentum direction of the initial electron inside a narrow cone with
opening angle $\theta_0\ll1$, and ii) the region with the photon emitted
along the scattered electron. Taking into account the identity
of photons with the statistical factor $\frac{1}{2!}$ we obtain
the cross--section:

\begin{eqnarray}
\dd\sigma_{SC}^{HH}&=& \frac{\alpha^4}{2\pi}\int
\frac{\dd^2\vecc{q}^{\bot}_2}{\pi((\vecc{q}^{\bot}_2)^2)^2}
\int \frac{\dd^2\vecc{q}^{\bot}_1}{\pi}
\int\limits_{x_c}^{1-2\Delta}\dd x \\ \nonumber
&\times& \int\limits_{\Delta}^{1-x-\Delta}
\frac{\dd x_1\dd x_2}{x_1x_2x}\delta(1-x_1-x_2-x)
\int R \frac{\dd^2\vecc{k}^{\bot}_1}{\pi}\, ,
\end{eqnarray}

where
\begin{eqnarray}
\int R \frac{\dd^2\vecc{k}^{\bot}_1}{\pi}
&=& 2(\vecc{q}^{\bot}_2)^2 Q_1^4\int\frac{\dd^2\vecc{k}^{\bot}_1}{\pi}
\biggl\{\frac{[1+(1-x_1)^2][x^2+(1-x_1)^2]}
{x_1(1-x_1)^2(2p_1k_1)(2p_1k_2)(2q_1k_2)}
\bigg|_{\vecc{k}_1\parallel\vecc{p}_1} \\ \nonumber
&+& \frac{x[1+(1-x_2)^2][x^2+(1-x_2)^2]}{x_1(1-x_2)^2
(2q_1k_1)(2p_1k_2)(2q_1k_2)}\bigg|_{\vecc{k}_1\parallel\vecc{q}_1}\biggr\}.
\end{eqnarray}

It is convenient to specify the kinematics: in the case of the emission
of the collinear photon with momentum $\vecc{k}_1$ parallel to
$\vecc{p}_1$ we have
\begin{eqnarray}
2p_1k_1&=&\frac{Q_1^2}{x_1}[(\vecc{k}^{\bot}_1)^2+\sigma^2x_1^2],\quad
2p_1k_2=\frac{Q_1^2}{x_2}(\vecc{k}^{\bot}_2)^2, \\ \nonumber
2q_1k_2&=&\frac{Q_1^2}{x_2x}
[x\vecc{q}^{\bot}_2-(1-x_1)\vecc{q}^{\bot}_1]^2, \qquad
\vecc{k}^{\bot}_2=-\vecc{q}^{\bot}_2-\vecc{q}^{\bot}_1\, ;
\end{eqnarray}
in the case when the photon is emitted along $\vecc{q}_1$ we have
\begin{eqnarray}
2k_1q_1=\frac{Q_1^2}{x_1x}[\sigma^2x_1^2
+ (x\vecc{k}^{\bot}_1-\vecc{q}^{\bot}_1)^2], \qquad
2p_1k_2=\frac{Q_1^2}{x_2}(\vecc{k}^{\bot}_2)^2,\\ \nonumber
2q_1k_2=\frac{Q_1^2}{x_2x}(\vecc{q}^{\bot}_1-x\vecc{q}^{\bot}_2)^2,\qquad
\vecc{k}^{\bot}_2=-\vecc{q}^{\bot}_2-\vecc{q}^{\bot}_1\frac{1-x_2}{x}\, ,
\end{eqnarray}
where $Q_1^2=\epsilon^2\theta_1^2$, $\sigma^2=m^2/Q_1^2$, and we introduced
two--dimensional vectors $\vecc{k}^{\bot}_2$, $\vecc{q}^{\bot}_1$ and
$\vecc{q}^{\bot}_2$ so that
$(\vecc{q}^{\bot}_1)^2=z_1$, $(\vecc{q}^{\bot}_2)^2=z_2$ and
$\widehat{\vecc{q}^{\bot}_1\vecc{q}^{\bot}_2}=\phi$.

The integration over $\dd^2\vecc{k}^{\bot}_1$ can be done with
single logarithmic accuracy:
\begin{eqnarray}
Q_1^2 \int\frac{\dd^2\vecc{k}^{\bot}_1}{\pi(2p_1k_1)}
\bigg|_{\vecc{k}_1\parallel\vecc{p}_1}=x_1L,
\quad
Q_1^2 \int\frac{\dd^2\vecc{k}^{\bot}_1}{\pi(2q_1k_1)}
\bigg|_{\vecc{k}_1\parallel\vecc{q}_1}=\frac{x_1}{x}L.
\end{eqnarray}

It is also necessary, here, to consider the kinematical restrictions on the
integration variables $\phi$ and $z_1$. When the photon is emitted
within an angle $\theta_0$ along the
direction of the momentum of the initial electron, $\theta_0$
represents the angular range to be filled by collinear kinematics events.
We assign to the semi--collinear kinematics the events characterized by

\begin{eqnarray}
i)\ \left|\frac{\vecc{k}^{\bot}_2}{x_2}\right|>\theta_0\, , \qquad
ii)\ \bigg|\frac{\vecc{q}^{\bot}_1}{x}
- \frac{\vecc{k}^{\bot}_2}{x_2}\bigg| > \theta_0,
\end{eqnarray}
where the region $i)$ the photon with four--momentum $k_2$ escapes
the narrow cone with opening angle $\theta_0$ along the momentum
direction of the initial electron. In the region $ii)$ the same happens
for the final electron.

We can rewrite the conditions above in terms of the variables $z_1$ and $\phi$
as follows:

\begin{eqnarray}
i)&& 1> \cos\phi > -1+\frac{\lambda^2-(\sqrt{z_1}-\sqrt{z_2})^2}
{2\sqrt{z_1z_2}},\qquad
|\sqrt{z_1}-\sqrt{z_2}|<\lambda, \nonumber \\ \nonumber
ii)&& 1>\cos\phi > -1,\qquad |z_1-z_2|>2\sqrt{z_2}\lambda,
\\ \nonumber
iii)&& 1>\cos\phi>-1+\frac{\frac{x^2}{(1-x_1)^2}\lambda^2-
(\sqrt{z_1}-\frac{x}{1-x_1}\sqrt{z_2})^2}{2\sqrt{z_1z_2}\frac{x}{1-x_1}},\\ \nonumber
&& \qquad |\sqrt{z_1}-\frac{x\sqrt{z_2}}{1-x_1}|<\lambda\frac{x}{1-x_1}\, ,
\\ \nonumber
iv)&& 1>\cos\phi>-1,\qquad |z_1-\frac{x^2}{(1-x_1)^2}z_2|>2\lambda\sqrt{z_2}
\frac{x^2}{(1-x_1)^2},
\end{eqnarray}
where $\lambda=x_2\theta_0/\theta_1$. In our calculation we take
the parameter $\lambda\ll 1$. Indeed, the restrictions on $\theta_0$
for collinear kinematics calculations
are $ \varepsilon\theta_0\gg m$ or $\theta_0\gg 10^{-5}$ at LEP energies.
On the other
hand the experimental conditions on $\theta_1$ are $\theta_1>10^{-2}$. Therefore
we can take $\lambda\ll 1$ within our accuracy.

Analogous considerations can be made for the case when a photon with
momentum $k_1$ is emitted along the direction of the final electron.
In regions $ii)$ and $iv)$ we may do the integration over the azimuthal
angle:
\begin{eqnarray}
\int\limits_{0}^{2\pi}\frac{\dd\phi}{2\pi(2p_1k_2)(2q_1k_2)}
\bigg|_{\vecc{k}_1\parallel\vecc{p}_1}
&=& \frac{x_2xQ_1^{-4}}{(1-x_1)z_1-xz_2}\biggl[\frac{1}{|z_2-z_1|}
- \frac{x(1-x_1)}{|x^2z_2-(1-x_1)^2z_1|}\biggr],\\
\int_{0}^{2\pi}\frac{\dd\phi}{2\pi(2p_1k_2)(2q_1k_2)}
\bigg|_{\vecc{k}_1\parallel\vecc{q}_1}
&=& \frac{x_2x^3(1-x_2)^{-2}Q_1^{-4}}{z_1-z_2x^2/(1-x_2)}\biggl[\frac{1}
{|z_1-z_2\frac{x^2}{(1-x_2)^2}|}-\frac{1-x_2}{|z_1-x^2z_2|}\biggr].
\end{eqnarray}

The integration of regions $i)$, $iii)$ has the form

\begin{eqnarray}
{\cal{I}}&=&\int\dd z_1\int\frac{\dd\phi}
{2\pi(z_1+z_2+2\sqrt{z_1z_2}\,\cos\phi)}
\bigg|_{|\sqrt{z_1}-\sqrt{z_2}|<\lambda} \\ \nonumber
&=& \frac{2}{\pi}\int \frac{\dd z}{|z_1-z_2|}
\arctan\biggl\{\frac{(\sqrt{z_1}-\sqrt{z_2})^2}{|z_1-z_2|}
\tan\frac{\phi_0}{2}\biggr\},
\end{eqnarray}
where
\begin{eqnarray}
\phi_0=\arccos\biggl(-1+\frac{\lambda^2
- (\sqrt{z_1}-\sqrt{z_2})^2}{2\sqrt{z_1z_2}}\biggr).
\end{eqnarray}

The result reads
\begin{eqnarray}
{\cal{I}}=2\;\ln2.
\end{eqnarray}

We give here the complete contribution of the semi--collinear region:
\begin{eqnarray*}
\dd\sigma_{\mbox{\tiny s-coll}}^{HH}&=&\frac{\alpha^2{\cal{L}}}{4\pi^2}
\int\limits_{x_c}^{1-2\Delta}\dd x
\int\limits_{\Delta}^{1-x-\Delta}\frac{\dd x_1 \dd x_2
\delta(1-x-x_1-x_2)}{x_1x_2(1-x_1)^2}[1+(1-x_1)^2][x^2+(1-x_1)^2] \\
 &\times& \int\limits_{1}^{\rho^2}\frac{\dd z}{z^2}
\biggl\{ \ln\frac{z\theta_1^2}{\theta_0^2}\bigl[1+\Theta(\rho^2x^2-z)
+ 2\Theta(\rho^2(1-x_1)^2-z)\bigr] \\
&+& \Theta(\rho^2x^2-z)\ln\frac{(z-x^2)(\rho^2x^2-z)}
{x^2(z-x(1-x_1))(\rho^2x(1-x_1)-z)} \\
&+& \Theta(z-\rho^2(1-x_1)^2)
\biggl[\ln\frac{(z-\rho^2(1-x_1)x)(z-(1-x_1)^2)}
{(\rho^2(1-x_1)^2-z)(z-x(1-x_1))} \\
&+& \ln\frac{(\rho^2(1-x_1)-z)(z-(1-x_1)^2)}
{(\rho^2(1-x_1)^2-z)(z-(1-x_1))}\biggr]+\Theta(z-\rho^2 x^2)
\ln\frac{z-\rho^2 x(1-x_1)(z-x^2)}{(\rho^2 x^2-z)(z-x(1-x_1))}
 \\
&+& \Theta(\rho^2(1-x_1)^2-z)
\biggl[\ln\frac{(z-(1-x_1)^2)(\rho^2(1-x_1)^2-z)}{(\rho^2x(1-x_1)-z)
(z-x(1-x_1))(1-x_1)^2} \\
&+& \ln\frac{(z-(1-x_1)^2)(\rho^2(1-x_1)^2-z)}
{(\rho^2(1-x_1)-z)(z-(1-x_1))(1-x_1)^2} \biggr]
+ \ln\frac{(z-1)(\rho^2-z)}{(z-(1-x_1))(\rho^2(1-x_1)-z)}
\biggr\}.
\end{eqnarray*}
To see the cancellation of the auxiliary parameter $\theta_0/\theta_1$
we give here the relevant part of the contribution for the collinear region :
\begin{eqnarray*}
\Sigma_{\mbox{\tiny coll}}^{HH}&=&\frac{\alpha^2}{4\pi^2}
\int\limits_{x_c}^{1-2\Delta}\dd x
\int\limits_{\Delta}^{1-x-\Delta}\frac{\dd x_1\; \dd x_2
\delta(1-x-x_1-x_2)}{x_1x_2(1-x_1)^2}[1+(1-x_1)^2][x^2+(1-x_1)^2]\\
&\times&\int\limits_{1}^{\rho^2}\frac{\dd z}{z^2}
\biggl(L^2+2L\ln\frac{\theta_0^2}{z\theta_1^2}\biggr)
\biggl[\frac{1}{2} + \frac{1}{2}\Theta(\rho^2x^2-z)
+ \Theta(\rho^2(1-x_1)^2-z) \biggr] + \dots \; .
\end{eqnarray*}
We see from the above expression that the dependence on $\theta_0/\theta_1$
disappears in the sum of the contributions for the
collinear and semi--collinear regions.
The total sum is given by Eq.~(\ref{eq54}).

%\newpage
%\end{document}

\newpage
\section*{Appendix C}
{\bf Virtual corrections to single photon
emission cross--section}
\vskip 20.0pt

\setcounter{equation}{0}
\renewcommand{\theequation}{C.\arabic{equation}}

The cross--section for single hard photon bremsstrahlung
containing virtual and real soft photon corrections
may be written as follows:

\begin{eqnarray}
\dd\sigma^{H(S+V)}=\frac {\alpha^3 \dd x \dd^2\vecc{q}^{\bot}_2
\dd^2\vecc{q}^{\bot}_1}{2\pi^2 x(1-x)
(\vecc{q}^{\bot}_2)^4}R,\qquad R=\lim_{(2p_1p_2)\to \infty}
\frac{4 p_{2\rho} p_{2\sigma} K_{\rho\sigma}}{(2p_1p_2)^2}\, .
\end{eqnarray}

We define the Compton tensor with a heavy photon as~\cite{r11}:
\begin{equation}
K_{\rho\sigma}=\frac{1}{8\pi\alpha}\sum\limits_{spins}M_{\rho}M_{\sigma}^*,
\end{equation}
where $M_{\rho}$ is the matrix elements of the Compton scattering process
\begin{eqnarray}
&& \gamma^*(q)\ +\ e(p_1)\longrightarrow \gamma(k)\ +\ e(p_2), \\ \nonumber
&& k^2=0,\qquad p_1^2=p_2^2=m^2,
\end{eqnarray}
$\rho$ is the polarization index of the heavy photon $(q^2\ne 0)$.
The Bose symmetry and the gauge-invariance requirements provide
the following general form of $K_{\rho\sigma}$:
\begin{eqnarray}
K_{\rho\sigma} &=& \frac{1}{2}(P_{\rho\sigma}+P^*_{\sigma\rho}), \\ \nonumber
P_{\rho\sigma} &=& \tilde g_{\rho\sigma}(B_g+\frac{\alpha}{2\pi}T_g)
+ \tilde p_{1\rho}\tilde p_{1\sigma}(B_{11}+\frac{\alpha}{2\pi}T_{11})
+ \tilde p_{2\rho}\tilde p_{2\sigma}(\tilde B_{11}+\frac{\alpha}{2\pi}
\tilde T_{11}) \\ \nonumber
&+& \tilde p_{1\rho}\tilde p_{2\sigma}(B_{12}+\frac{\alpha}{2\pi}T_{12})
+ \tilde p_{2\rho}\tilde p_{1\sigma}(\tilde B_{12}+\frac{\alpha}{2\pi}
\tilde T_{12}), \\ \nonumber
q_{\rho}P_{\rho\sigma} &=& 0,\qquad q_{\sigma}P_{\rho\sigma}=0,
\end{eqnarray}
where
\begin{eqnarray}
\tilde{g}_{\rho\sigma}=g_{\rho\sigma}-\frac{q_{\rho}q_{\sigma}}{q^2}\, ,
\quad \tilde{p}_{1\rho}=p_{1\rho}-q_{\rho}\frac{p_1q}{q^2}\, , \quad
\tilde{p}_{2\rho}=p_{2\rho}-q_{\rho}\frac{p_2q}{q^2}\, .
\end{eqnarray}
Coefficients $B_i$ and $T_i$ before different tensor structures
depend on the kinematical invariants
\begin{eqnarray}
s=(p_2+k)^2-m^2,\quad t^2=(p_1-k)^2-m^2,\quad u=(p_1-p_2)^2,\quad
s+t+u=q^2.
\end{eqnarray}
In the Born approximation we have
\begin{eqnarray}
B_g &=& \frac{1}{st}[(s+u)^2+(t+u)^2]-2m^2q^2\left(\frac{1}{s^2}
+ \frac{1}{t^2}\right),\qquad B_{12} = 0, \nonumber \\
B_{11} &=& \frac{4q^2}{st}-\frac{8m^2}{s^2}\, \qquad
B_i=B_i(s,t,u),\qquad \tilde{B}_i=B_i(t,s,u),\quad i=g,\, 12\,, 11.
\end{eqnarray}
Contributions in the one--loop approximation are:
\begin{eqnarray}
T_g &=& r B_g + \Biggl\{-\frac{1}{s}\biggl(\frac{u}{s}q^2
+ \frac{1}{t}(2ub+s^2)\biggr)G
+ u\biggl(\frac{q^2}{st}-\frac{2}{a}\biggr)(l_q-l_u) \\ \nonumber
&+& \frac{c}{s}\biggl(\frac{3u}{b}-1\biggr)(l_q-l_t)
+ \frac{u^2-s^2}{2st} + \frac{m^2}{t^2}\biggl[\biggl(
\frac{t}{\bar{t}}(4b-u) - 2s\biggr)l_t \\ \nonumber
&-& 2u + 4bG - 2q^2(l_q^2-l_u^2) + 2\biggl(b+\frac{m^2s}{t}\biggr)F
- bn \biggr] + (t\leftrightarrow s+i0) \Biggr\}, \\ \nonumber
T_{12} &=& \frac{2}{st}\Biggl\{\frac{q^2}{s^2}c(u-s)G
+ \frac{q^2}{t^2}(uq^2-st)\tilde{G}
- 2q^2\biggl(\frac{uq^2}{st}+ \frac{2u-s+t}{a}\biggr)(l_q-l_u) \\ \nonumber
&-& \frac{4}{a}(u^2-cs)\biggl(\frac{q^2}{a}(l_q-l_u)-1\biggr)
+ \frac{q^2}{c^2}(2c+t)\biggl(s-\frac{u}{t}q^2\biggr)(l_q-l_s) \\ \nonumber
&-& \frac{q^2c}{bs}(2u-s)(l_q-l_t) + 8u + 3t - s + \frac{2us}{c} \\ \nonumber
&+& m^2\biggl[\frac{2}{\bar{t}t}(ut+2m^2b)l_t
+ 2\biggl(\frac{t}{\bar{s}}+\frac{2c}{s}\biggr)l_s - \frac{u+2b}{t}
- \frac{u+4c}{s} + \frac{2c}{s_a}\tilde{N} + Q \\ \nonumber
&+& \frac{b}{u}R + \frac{c}{u}\tilde{R} + \frac{s+b}{t}n
- \frac{u}{s}\tilde{n} \biggr] \Biggr\}, \\ \nonumber
T_{11} &=& rB_{11} + \frac{2}{st}\Biggl\{
- q^2\biggl(1+\frac{u^2}{s^2}\biggr)G
- q^2\biggl(2+\frac{b^2}{t^2}\biggr)\tilde{G}
+ 2q^2\biggl(\frac{b^2}{st}+\frac{2u}{a}\biggr)(l_q-l_u) \\ \nonumber
&+& \frac{4}{a}(u^2-bt)\biggl(\frac{q^2}{a}(l_q-l_u)-1\biggr)
+ \frac{q^2b^2}{c^2t}(2c+t)(l_q-l_s)
+ \frac{q^2}{s}(2u-s)(l_q-l_t) \\ \nonumber
&-& 4u - 2q^2 + t - \frac{b^2}{c}
+ m^2\biggl[ - \frac{b}{t}(4l_t-5)
+ 2\biggl(\frac{(c+t)^2}{c\bar{s}}-\frac{2c}{s}\biggl)l_s \\ \nonumber
&+& \frac{u}{s}\biggl(2+\frac{u}{c}\biggr)
+ \frac{8t}{s}\biggl(\tilde{G} + \tilde{F} - \frac{1}{2}l_q^2
+ \frac{1}{2}l_u^2\biggr) \\ \nonumber
&-& \frac{2u}{s_a}\tilde{N} - Q - \frac{b^2}{u^2}R - \tilde{R}
+ \frac{b}{t}n - \frac{1}{cs}(2c^2-u^2)\tilde{n}\biggr]\Biggr\}, \\ \nonumber
r &=& 4(l_u-1)\ln\frac{\lambda}{m} - l_u^2 + 3l_q + \frac{\pi^2}{3}
- \frac{9}{2}\, ,
\end{eqnarray}
where the following notation is used:
\begin{eqnarray}
\bar{t} &=& t+m^2,\quad \bar{s}=s+m^2,\quad a=s+t,\quad b=s+u,\\ \nonumber
c &=& t+u,\quad s_a=s-\frac{m^2}{u}a,\quad t_a=t-\frac{m^2}{u}a,\\ \nonumber
G &=& (l_q-l_u)(l_q+l_u-2l_t)-2\mbox{Li}_2(1)
- 2\mbox{Li}_2\left(1-\frac{q^2}{u}\right)
+ 2\mbox{Li}_2\left(1-\frac{t}{q^2}\right), \\ \nonumber
F &=& - \mbox{Li}_2\left(1+\frac{t}{m^2}\right) + \mbox{Li}_2(1),\quad
N = 2l_t(l_q-l_u) + 2\mbox{Li}_2\left(1-\frac{q^2}{u}\right) - 2F,
\\ \nonumber
Q &=& \frac{4c}{s}\left(1-\frac{m^2}{s}\right)\tilde{F}
- 4b\frac{m^2}{t^2}F,\quad
n = \frac{m^2}{\bar{t}}\left(\frac{t}{\bar{t}}l_t-1\right), \\ \nonumber
R &=& \frac{u}{t_a}\left(1+\frac{m^2}{t_a}\right)N
- \frac{2u^2}{t_a}\biggl[\frac{1}{a}(l_q-l_u)-\frac{1}{q^2}l_t\biggr],
\\ \nonumber
l_t &=& \ln\left(-\frac{t}{m^2}\right),\quad
l_q=\ln\left(-\frac{q^2}{m^2}\right),\quad
l_u=\ln\left(-\frac{u}{m^2}\right),\quad
l_s=\ln\left(\frac{-s-i0}{m^2}\right).
\end{eqnarray}

In order to avoid the dependence of the Compton tensor on the photon mass
$\lambda$ we need to take into account the emission of an additiona soft 
photon with energy $\omega \leq \Delta \varepsilon $. The 
corresponding contribution, which has to be added to the quantity $r$, in the
case of small angle Bhabha scattering reads [15]:

\begin{eqnarray}
\rho=2(l_u-1)(2\ln\frac{m}{\lambda}+2\ln\Delta-\ln x)+l_u^2-\ln^2 x
-\frac{\pi^2}{3}.
\end{eqnarray}

Note that in the case of small--angle Bhabha scattering the contributions
of the $\tilde{g}_{\rho\sigma}$ tensor structure is suppressed
by factor $\theta^2$ relatively to terms containing
$\tilde{p}_{i\rho}\tilde{p}_{j\sigma}$.

We need only limited values of $T_{ik}$ in the cases of $s_1\ll|t_1|$ and
$|t_1|\ll s_1$ at fixed $q^2$ and $u_1=-2p_1q_1$.
In the case of small $s_1$ we have
$s_1\equiv s=[m^2(1-x)^2+(\vecc{q}^{\bot}_2x
+ \vecc{q}^{\bot}_1)^2]/[x(1-x)]$
(we omit in the remaining part of this Appendix the
subscript $1$ in the notation of invariants of the Compton subprocess).

Taking into account that, at small $s$,  ~$q^2=-(\vecc{q}^{\bot}_2)^2$,
$t=-(1-x)(\vecc{q}^{\bot}_2)^2$ and $u=-(\vecc{q}^{\bot}_2)^2x$, we
derive the following expressions
for $h=\rho+r$ and $T$ in this limit:
\begin{eqnarray}
h_{s\ll|t|}&=&2(L-1+\ln x)(2\ln\Delta-\ln x) + 3L - \ln^2x - \frac{9}{2}\, ,
\\ \nonumber
T_{s\ll|t|}&=&\frac {2}{s(1-x)}\biggl\{4(1+x^2)\biggl[\ln x\;
\ln\frac{(\vecc{q}^{\bot}_2)^2}{s}
- \mbox{Li}_2(1-x) \biggr] \\ \nonumber
&-& 1 + 2x +x^2 \biggr\} - \frac{16m^2}{s^2}\ln x L.
\end{eqnarray}

In the case of small $|t|$ we have:
\begin{eqnarray}
h_{|t|\ll s}&=&2(L-1-\ln x)(2\ln\Delta-\ln x) + 3L - \ln^2x
- \frac{9}{2}\, , \\ \nonumber
T_{|t|\ll s}&=&\frac{2x}{t(1-x)}\biggl\{4(1+x^2)\biggl[\ln x\;
\ln\frac{(\vecc{q}^{\bot}_2)^2}{-t}
- \frac{1}{2}\ln^2x - \mbox{Li}_2(1-x) \biggr] \\ \nonumber
&-& 1 - 2x + x^2\biggr\} + \frac{16m^2x^2}{t^2}\ln x\, L.
\end{eqnarray}
One can be convinced that all terms reinforced by large logarithms
($L$ or $L^2$) arise only from these regions. We omit here only terms
of order $(\alpha/\pi)^2$.

The further integration is straightforward. We show here the most
important moments. The contribution
of the $h$ containing terms gives
(in close analogy with the Born contribution):

\begin{eqnarray}\label{c6}
\Sigma_{h}^{H(S+V)}&=&\frac{1}{2}\bigl(\frac{\alpha}{\pi}\bigr)^2
\int\limits_1^{\rho^2}
\frac{\dd z}{z^2}L\int\limits_{x_c}^{1-\Delta}\dd x\frac{1+x^2}{1-x}\biggl\{
(1+\Theta(\rho^2x^2-z))\\ \nonumber
&\times& \biggl[L(2\ln\Delta-\ln x+\frac{3}{2})
+ (2\ln\Delta-\ln x)(\ln x-2)-\frac{1}{2}\ln^2x-\frac{15}{4}\biggr]\\ \nonumber
&+& (2\ln\Delta-\ln x+\frac{3}{2})\;k(x,z)
- 2\ln x(2\ln\Delta-\ln x)\Theta(\rho^2x^2-z)\biggr\}.
\end{eqnarray}

To obtain the contributions from $T$ we consider at first new types
of integrals:
\begin{eqnarray}
I_{s\{t\}}&=&Q_1^2\int\frac{\dd^2\vecc{q}_2^{\bot}}
{\pi(\vecc{q}_2^{\bot})^4}
\int\frac{\dd^2\vecc{q}_1^{\bot}}{\pi s\{t\}}
\ln\frac{(\vecc{q}_2^{\bot})^2}{s\{-t\}}\, ,
\\ \nonumber
i_{s\{t\}}&=&Q_1^2\int \frac{\dd^2\vecc{q}_2^{\bot}}
{\pi(\vecc{q}_2^{\bot})^4}
\int\frac{\dd^2\vecc{q}_1^{\bot}}{\pi s\{t\}}\, ,  \quad
%\\ \nonumber
m_{s\{t\}}=Q_1^2\int \frac{\dd^2\vecc{q}_2^{\bot}}
{\pi(\vecc{q}_2^{\bot})^4}
\int\frac{\dd^2\vecc{q}_1^{\bot}m^2}{\pi s^2\{t^2\}}\, .
\end{eqnarray}

Denoting $\sigma^2(1-x)^2+(\vecc{q}^{\bot}_2x-\vecc{q}_1^{\bot})^2/Q_1^2$ as
$a+b\cos\phi$ and using the expressions
\begin{eqnarray}
\frac{1}{2\pi}\int\limits_0^{2\pi}\frac{\dd\phi}{a+b\cos\phi}
&=&\frac{1}{\sqrt{a^2-b^2}}\, ,
\\ \nonumber
\frac{1}{2\pi}\int\limits_0^{2\pi}\dd\phi
\frac{\ln(a+b\cos\phi)}{a+b\cos\phi}
&=&\frac{1}{\sqrt{a^2-b^2}}
\ln\frac{2(a^2-b^2)}{a+\sqrt{a^2-b^2}}
\end{eqnarray}
with
$a^2-b^2=(z_1-x^2z_2)^2+4 \sigma^2(1-x)^2 x^2 z_2\, $,
~$\sigma^2= m^2/Q_1^2$ and
$z_{1,2}=(\vecc{q}_{1,2}^{\bot})^2/Q_1^2$ we derive that
\begin{eqnarray}
I_s&=&(1-x)x\int\limits_1^{\rho^2}\frac{\dd z_2}{z_2^2}
\int\limits_{x^2}^{x^2\rho^2}\dd z_1 \\ \nonumber &\times&
\frac{\ln(z_2^2(1-x)x^3)-\ln[(z_1-z_2x^2)^2+4\sigma^2x^2(1-x)^2z_2]}
{\sqrt{(z_1-x^2z_2)^2+4\sigma^2x^2(1-x)^2z_2}}\, .
\end{eqnarray}
Since we are evaluating with logarithmic accuracy, we may consider the
contribution of the region $|z_1-x^2z_2|<\eta$,
$\sigma^2 \ll \eta \ll 1$
when integrating over $z_1$. The result reads
\begin{eqnarray}\label{c12}
I_s=x(1-x)\int\limits_1^{\rho^2}\frac{\dd z_2}{z_2^2}L
\biggl[\frac{1}{2}L+\ln\frac{x}{1-x}\biggr].
\end{eqnarray}

The rest of the integrals can be calculated in the same way, and we have:
\begin{eqnarray}\label{c13}
I_t&=&-(1-x)\int\limits_1^{\rho^2x^2}\frac{\dd z_2}{z_2^2}L
\biggl[\frac{1}{2}L + \ln\frac{1}{1-x}\biggr], \quad
i_t=-(1-x)\int\limits_1^{\rho^2x^2}\frac{\dd z_2}{z_2^2}L,\\ \nonumber
i_s&=&x(1-x)\int\limits_1^{\rho^2}\frac{\dd z_2}{z_2^2}L, \quad
m_t=\int\limits_1^{\rho^2x^2}\frac{\dd z_2}{z_2^2}\, ,\quad
m_s=x^2\int\limits_1^{\rho^2x^2}\frac{\dd z_2}{z_2^2}\, .
\end{eqnarray}
Using (\ref{c12}) and (\ref{c13}) we may represent the final result for
the contribution to $\Sigma^{H(S+V)}$ due to the $T$ term as
\begin{eqnarray}\label{c14}
\Sigma^{H(S+V)}_{T}&=&\frac{1}{2}\bigl(\frac{\alpha}{\pi}\bigr)^2
\int\limits_1^{\rho^2}\frac{\dd z}{z^2}\;L\;
\int\limits_{x_c}^{1-\Delta}\dd x\frac{1+x^2}{1-x}\biggl\{
\biggl(\frac{1}{2}L+\ln\frac{x}{1-x}\biggr)\ln x
\\ \nonumber
&+& \zeta_2 - \mbox{Li}_2(x) + \frac{x^2+2x-1}{4(1+x^2)}
-\frac{2x\ln x}{1+x^2}
-\Theta(x^2\rho^2-z)\biggl[\biggl(\frac{1}{2}L+\ln\frac{1}{1-x}\biggr)\ln x
\\ \nonumber
&-& \frac{1}{2}\ln^2x
-\mbox{Li}_2(1-x) - \frac{1+2x-x^2}{4(1+x^2)}
- \frac{2x\ln x}{1+x^2}\biggr]\biggr\}.
\end{eqnarray}

The total contribution to $\Sigma^{H(S+V)}$ (one-side hard photon
emission with virtual and soft photon corrections) is the sum of
(\ref{c6}) and (\ref{c14}):
\begin{eqnarray}
\Sigma^{H(S+V)}=\Sigma^{H(S+V)}_{\rho}+\Sigma^{H(S+V)}_{T}.
\end{eqnarray}
This quantity is given in Eq.~(\ref{eq49}).

%\newpage
%\end{document}

%\chapter{}
\newpage
\section*{Appendix D}
{\bf Leading logarithmic contribution to $\Sigma^{\gamma\gamma}$}
\vskip 20.0pt
\setcounter{equation}{0}
\renewcommand{\theequation}{D.\arabic{equation}}

Here we  show that the main logarithmic terms can be summed
according to the renormalization group. The sum, as given in the text,
may be written as:

\begin{eqnarray} \label{d1}
S &=& 2\bigl(\frac{\alpha}{\pi}\bigr)^2
\int\limits_1^{\rho^2} \frac{\dd z}{z^2}L^2
\biggl\{ \int\limits_{x_c}^1\dd x\, \delta(1-x)
\biggl(\ln^2\Delta + \frac{3}{2}\ln\Delta + \frac{9}{16}\biggr) \\ \nonumber
&\times& (1+ \Theta(x^2\rho^2-z))
+\frac{1}{2}\int_{x_c}^{1-\Delta}
\frac{1+x^2}{1-x}(2\ln \Delta-\ln x+ \frac{3}{2})(1+ \Theta(x^2\rho^2-z))
\\ \nonumber
&+& \frac{1}{4} \int\limits_{x_c}^{1-2\Delta}\dd x\biggl[2\frac{1+x^2}{1-x}
\ln\frac{1-x-\Delta}{\Delta}
+ \frac{1}{2}(1+x)\ln x-1+x\biggr][1+3\Theta(x^2\rho^2-z)] \\ \nonumber
&+& \frac{1}{4}\int\limits_{x_c}^{1}\dd x\biggl[
2\frac{1+x^2}{1-x}\ln\left(\frac{(1-x-\Delta)(\rho-\sqrt{z})}
{\Delta(\sqrt{z}-\rho x)}\,\sqrt{x}\right)+x-1 \\ \nonumber
&-& \frac{1}{2}(1+x)\ln\frac{\rho^2}{z}+
\frac{\sqrt{z}}{\rho}-\frac{x\rho}{\sqrt{z}}\biggr]
\Theta(z-x^2\rho^2)\biggr\}.
\end{eqnarray}
One can see that the dependence on the parameter $\Delta$ disappears in
the expression above.

We will now show that eq.~(\ref{d1}) is equivalent to eq.~(\ref{eq58}).
Let us transform eq.~(\ref{eq58}) using the substitution
\begin{eqnarray} \label{d2}
\Theta(t^2\rho^2-z)=\frac{1}{2}(1+\Theta(x^2\rho^2-z))
+\frac{1}{2}\Theta(z-x^2\rho^2) - \Theta(z-t^2\rho^2),
\end{eqnarray}
and changing the order of integration in the last term:
\begin{eqnarray} \label{d3}
\int\limits_{x}^{1}\dd t \int\limits_{\rho^2t^2}^{\rho^2}\dd z
&=& \int\limits_{\rho^2x^2}^{\rho^2}\dd z \int\limits_{x}^{\sqrt{z}/\rho}\dd t
%\\ \nonumber &=&
=\int\limits_{1}^{\rho^2}\dd z \Theta(z-\rho^2x^2)
\int\limits_{x}^{\sqrt{z}/\rho}\dd t.
\end{eqnarray}

By evaluating the integral over $t$, and using the explicit
expressions for the splitting functions one can verify the coincidence
of Eqs.~(\ref{d1}) and (\ref{eq58}). In an analogous way one can prove
the validity of the representation (\ref{eq60}) for $\Sigma_{\gamma}^{\gamma}$.

Using the representation in Eq.~(\ref{eq59}) for the function $P^{(2)}$ one
can see that the above expression is equivalent to Eq.~(\ref{d1}).
In an analogous way one can prove the validity of representation
in Eq.~(\ref{eq60}) for $\Sigma^{\gamma}_{\gamma}$.

%\newpage
%\end{document}

%\chapter{}
\newpage
\section*{Appendix E}
{\bf Cancellation of the $\Delta$ dependence in the nonleading
contributions to $\Sigma^{2\gamma}$}
\vskip 20.0pt
\setcounter{equation}{0}
\renewcommand{\theequation}{E.\arabic{equation}}

Let us consider the singular
nonleading terms in $\Sigma^{2\gamma}$
in the limiting case $\Delta\to 0$. Dropping the common
factor $(\alpha/\pi)^2{\cal L}\int\dd z/z^2$, we give below
the various contributions separately.

Let us consider first $\Sigma^{\gamma\gamma}$.
The contributions from
the soft photon radiation and virtual corrections are:

\begin{eqnarray} \label{e1}
(\Sigma^{VV+VS+SS})_\Delta=\ln\Delta(-7-4\ln\Delta)(1,\rho^2),
\end{eqnarray}
where we denote by $(a,b)$ the limits of the integration over $z$:
$(a,b)=\Theta(z-a)\Theta(b-z)$.

The contribution due to the virtual
corrections to the single hard photon emission
gives\footnote{In our contribution to the Tenessee--94 workshop
(see [2]) there are some misprints, which are corrected here}

\begin{eqnarray} \label{e2}
(\Sigma^{H(S+V)})_\Delta&=&\frac{1}{2}\ln\Delta\biggl\{
(1,\rho^2)(16\ln\Delta+14)+ 4\int\limits_{\tilde{x}_c}^{1}
\dd x\;\frac{1+x^2}{1-x}[(1,\rho^2)-(1,\rho^2x^2)] \\ \nonumber
&+& 4[-2\ln(1-x_c)-2\ln(1-\tilde{x}_c) + \int\limits_{x_c}^{1}\dd x\;(1+x)
+ \int\limits_{\tilde{x}_c}^{1}\dd x\;(1+x)] \\ \nonumber
&+& 2 \int\limits_{x_c}^{1}\dd x\;\frac{1+x^2}{1-x}
[(1,\rho^2)-(1,\rho^2x^2)]\ln x + 2\int\limits_{x_c}^{1}\dd x\;
\frac{1+x^2}{1-x} \;k(x,z) \biggr\},
\end{eqnarray}
where $\tilde{x}_c=\mbox{max}(x_c,1/\rho)$ and the quantity $k(x,z)$
is defined in Eq.~(\ref{eq30}).
The singular part in the contribution in Eq.~(\ref{eq54}) due to double hard
photon bremsstrahlung reads:

\begin{eqnarray} \label{e3}
(\Sigma^{HH})_{\Delta}&=&\ln\Delta\biggl\{\int\limits_{x_c}^{1}\dd x\;
\frac{1+x^2}{1-x}[-(1,\rho^2)L_1 - (1,\rho^2x^2)L_2
- ((1,\rho^2)-(1,\rho^2x^2))L_3] \\ \nonumber
&-& \int\limits_{x_c}^{1}\dd x\;(3+x)
- \int\limits_{\tilde{x}_c}^{1}\dd x\;(3+x) - 4\ln\Delta
+ 4\ln(1-x_c) + 4\ln(1-\tilde{x}_c) \\ \nonumber
&-& \int\limits_{\tilde{x}_c}^{1}\dd x\;\frac{1+x^2}{1-x}[(1,\rho^2)
- (1,\rho^2x^2)] \biggr\}.
\end{eqnarray}
It is possible to verify the cancellation:
\begin{eqnarray} \label{e4}
(\Sigma^{VV+VS+SS})_\Delta + (\Sigma^{H(S+V)})_\Delta
+ (\Sigma^{HH})_{\Delta} = 0.
\end{eqnarray}

The corresponding contributions to $\Sigma^{\gamma}_{\gamma}$ are:

\begin{eqnarray} \label{e5}
(\Sigma^{S+V}_{S+V})_\Delta&=&\ln\Delta(-14-8\ln\Delta)(1,\rho^2), \\ \nonumber
(\Sigma^{H}_{S+V}+\Sigma_{H}^{S+V})_\Delta&=&\ln\Delta\biggl\{
2\int\limits_{x_c}^{1}\dd x\;\frac{1+x^2}{1-x}\; k(x,z)
+ (1,\rho^2)[16\ln\Delta + 14 - 8\ln(1-x_c) \\ \nonumber
&-& 8\ln(1-\tilde{x}_c) + 4\int\limits_{x_c}^{1}\dd x\;(1+x)
+ 4\int\limits_{\tilde{x}_c}^{1}\dd x\;(1+x)]  \\ \nonumber
&+& 4\int\limits_{\tilde{x}_c}^{1}\dd x\;\frac{1+x^2}{1-x}
[(1,\rho^2)-(1,\rho^2x^2)] \biggr\}, \\ \nonumber
(\Sigma^{H}_{H})_{\Delta}&=&\ln\Delta\biggl\{ - 8(1,\rho^2)
[\ln\Delta - \ln(1-x_c) - \ln(1-\tilde{x}_c)] \\ \nonumber
&-& 8\int\limits_{\tilde{x}_c}^{1}\dd x\;\frac{
(1,\rho^2)-(1,\rho^2x^2)}{1-x}
- 2\int\limits_{x_c}^{1}\dd x\;k(x,z)\frac{1+x^2}{1-x}
\\ \nonumber
&-& 4\int\limits_{x_c}^{1}\dd x\;(1+x)[(1,\rho^2)+(1,\rho^2x^2)]
\biggr\}.
\end{eqnarray}

Rearranging the last term in $(\Sigma^{H}_{H})_{\Delta}$ as
\begin{eqnarray} \label{e6}
&& - 4\int\limits_{x_c}^{1}\dd x\;(1+x)[(1,\rho^2)+(1,\rho^2x^2)] =
- 4\int\limits_{x_c}^{1}\dd x\;(1+x)(1,\rho^2) \\ \nonumber
&& \qquad - 4\int\limits_{\tilde{x}_c}^{1}\dd x\;(1+x)(1,\rho^2)
+ 4\int\limits_{\tilde{x}_c}^{1}\dd x\;(1+x)[(1,\rho^2)-(1,\rho^2x^2)],
\end{eqnarray}
we can see again the cancellation of the $\Delta$-dependence in the sum:
\begin{eqnarray} \label{e7}
(\Sigma^{S+V}_{S+V})_\Delta + (\Sigma^{H}_{S+V}+\Sigma_{H}^{S+V})_\Delta
+ (\Sigma^{H}_{H})_{\Delta} = 0.
\end{eqnarray}

%\newpage
%\end{document}

%\chapter{}
\newpage
\section*{Appendix F}
{\bf Relevant integrals for collinear pair production}

\vskip 20.0pt

\setcounter{equation}{0}
\renewcommand{\theequation}{F.\arabic{equation}}

We give here a list of the relevant integrals,
calculated within the logarithmic accuracy, for the collinear kinematical
region of hard pair production.

We use the definitions
in Eq.~(\ref{p6}) and we imply, in the left-hand side of the
expressions below, the general operation:
\begin{eqnarray} \label{f1}
\langle (\dots) \rangle \equiv \int\limits_{0}^{z_0}\!\!\!\; \dd z_1
\int\limits_{0}^{z_0}\!\!\!\; \dd z_2 \int\limits_{0}^{2\pi}\!\!\!\;
\frac{\dd \phi}{2\pi}\, (\dots)\, ,
\end{eqnarray}
with the conditions
$z_0=(\varepsilon \theta_0/m)^2 \gg 1$, $\ L_0=\ln z_0 \gg 1$.
The details of the calculations can be found in the Appendix of
Ref.~\cite{r13}. The results are:
\begin{eqnarray} \label{f2}
&& \langle \left(\frac{x_2D+(1-x_2)A}{DC} \right)^2 \rangle =
\frac{L_0}{(1-x_2)^2} \biggl\{ L_0 + 2\ln\frac{x_1x_2}{x} - 8 \\ \nonumber
&& \qquad + \frac{(1-x)^2(1-x_2)^2}{xx_1x_2} \biggr\}, \qquad
\langle \frac{1}{DC} \rangle = \frac{L_0}{x_1x_2(1-x_2)}
\bigl[\frac{1}{2}L_0 + \ln\frac{x_1x_2}{x} \bigr]\, , \\ \nonumber
&& \langle \left(\frac{x_2A_1-x_1A_2}{AD} \right)^2 \rangle =
\frac{L_0}{(1-x)^2} \biggl\{ L_0 + 2\ln\frac{x_1x_2}{x} - 8
+ \frac{(1-x)^2}{xx_1x_2} - \frac{4(1-x)}{x} \biggr\}\, , \\ \nonumber
&& \langle \frac{x_1A_2-x_2A_1}{AD^2} \rangle =
\frac{(x_1-x_2)L_0}{xx_1x_2(1-x)}\, , \qquad
\langle \frac{1}{A^2D} \rangle = \frac{-L_0}{(1-x)^3}, \\  \nonumber
&& \langle \frac{1}{AD} \rangle = \frac{-L_0}{x_1x_2(1-x)}
\bigl[\frac{1}{2}L_0 + \ln\frac{x_1x_2}{x} \bigr], \qquad
\langle \frac{1}{C^2D} \rangle = \frac{-L_0}{x_1(1-x_2)^3}\, , \\ \nonumber
&& \langle \frac{1}{AC} \rangle = \frac{-L_0}{x_1x_2^2}
\bigl[L_0 + 2\ln\frac{x_1x_2}{x} + 2\ln\frac{xx_2}{(1-x)(1-x_2)}
\bigr], \qquad
\langle \frac{1}{D^2} \rangle =\frac{L_0}{xx_1x_2}\, , \\ \nonumber
&& \langle \frac{A}{C^2D^2} \rangle = \frac{x_2L_0}{x_1(1-x_2)^4}, \qquad
\langle \frac{C}{A^2D^2} \rangle = \frac{-x_2L_0}{(1-x)^4}\, , \\ \nonumber
&& \langle \frac{A}{CD^2} \rangle = \frac{-L_0}{x_1(1-x_2)^2}
\bigl[ \frac{1}{2}L_0 + \ln\frac{x_1x_2}{x} \bigr]
+ L_0\frac{x_2x-x_1}{xx_1x_2(1-x_2)^2}\, , \\  \nonumber
&& \langle \frac{C}{AD^2} \rangle = \frac{-L_0}{x_1(1-x)^2}
\bigl[ \frac{1}{2}L_0 + \ln\frac{x_1x_2}{x} \bigr]
- L_0\left( \frac{x_1-x_2}{x_1x_2(1-x)^2} + \frac{1}{xx_2(1-x)} \right)\, .
\end{eqnarray}


\begin{thebibliography}{99}



\bibitem{r1}
     G. Altarelli, Lectures given at the E. Majorana Summer Scool, Erice,
       Italy, July 1993, CERN-TH.7072/93.

\bibitem{r1a}
LEP Electroweak Working Group, A Combination of Preliminary LEP
Electroweak Results from the 1995 Summer Conferences, 1995, CERN
report LEPEWWG/95--02.
\\
B. Pietrzyk, High Precision Measurements of the Luminosity at LEP,
Proceedings of the "Tennessee International Symposium on Radiative
Corrections: Status and Outlook", Gatlinburg, Tennessee, 27 June--
1 July, 1994, Ed. B.F.L. Ward (World Scientific, Singapore 1995).


\bibitem{r111}
       $Neutrino\;Counting$ in $Z\;Physics\;at\;LEP$,
       G. Barbiellini et al., L. Trentadue (conv.),
       G. Altarelli, R. Kleiss and C. Verzegnassi eds., CERN Report 89-08
 (1989).

\bibitem{r2}
       W. Beenakker, F.A.~Berends and S.C.~van~der~Marck, Nucl. Phys. B355
       (1991) 281;
\\     M. Cacciari, A. Deandrea, G. Montagna, O. Nicrosini and L. Trentadue,
 Phys.
       Lett. B271 (1991) 431;
\\     M. Caffo, H. Czyz and E. Remiddi, Nuovo Cim. 105A (1992) 277;
\\     K.S. Bjorkenvoll, G. F\"aldt and P. Osland, Nucl. Phys. B386 (1992)
       280, 303;
\\
       W.~Beenakker and B.~Pietrzyk, Phys. Lett. B296 (1992) 241; and
       B304 (1993) 366.
\\
       M. Cacciari, G. Montagna, O. Nicrosini and F. Piccinini, Comput.
       Phys. Commun. 90 (1995) 301. 


\bibitem{r3}
       S.~Jadach et al.,  Phys. Rev. D47 (1993) 3733;
\\     S.~Jadach, E.~Richter-Was, B.F.L.~Ward and
       Z.~Was, Phys. Lett. B260 (1991) 438, and 268 (1991) 253;
\\     S.~Jadach, M.~Skrzypek and B.F.L.~Ward, Phys. Lett. B257 (1991) 173.
\\     S.~Jadach et al., Phys. Lett. B353 (1995) 362.


\bibitem{r4}
R. Budny,  Phys. Lett. 55B (1975) 227; \\
D. Bardin, W. Hollik and T. Riemann, Z. Phys. C49 (1991) 485; \\
M. Boehm, A. Denner and W. Hollik, Nucl. Phys. B304 (1988) 687. \\

\bibitem{rx}
V.S.~Fadin, E.A.~Kuraev, L.N.~Lipatov, N.P.~Merenkov
and L.~Trentadue, Yad. Fiz. 56 (1993) 145; \\
S.~Jadach et al., Phys. Lett. B253 (1991) 469.

\bibitem{r33} V.S. Fadin, E.A. Kuraev, L.N. Lipatov, N.P. Merenkov and L. Trentadue,
Yad. Fiz. 56 (1993) 145.\\
V.S. Fadin, E.A. Kuraev, L.N. Lipatov, N.P. Merenkov and L. Trentadue,
Proceedings of the XXIX Rencontres de Moriond, M\`eribel, March 1994.\\
V.S. Fadin, E.A. Kuraev, L.N. Lipatov, N.P. Merenkov and L. Trentadue,
same Proceedings as in Ref. \cite{r1a} p.168.

\bibitem{r5}
R.V. Polovin, JEPT 31 (1956) 449 ; \\
F.A. Redhead, Proc. Roy. Soc. 220 (1953) 219; \\
F.A. Berends et al., Nucl. Phys. B68 (1974) 541.

\bibitem{r6}
S.~Eidelman, F.~Egerlehner, Z. Phys. C67 (1995) 585.

\bibitem{r7}
E.A. Kuraev, L.N. Lipatov and N.P.~Merenkov, Phys Lett. 47B (1973) 33;
preprint 46 LNPI, 1973; \\
H. Cheng, T. T. Wu,
Phys. Rev. 187 (1969) 1868; \\
V.S. Fadin, E.A. Kuraev, L.N. Lipatov, N.P. Merenkov and L. Trentadue,
Yad. Fiz. 56 (1993) 145.

\bibitem{r8} G. F\"aldt and P.~Osland, Nucl. Phys. B413 (1994) 16; Erratum
 ibidem B419 (1994) 404.

\bibitem{YFS}
D.R.~Yennie, S.C.~Frautchi, H.~Suura,
Ann. Phys. 13 (1961) 379.

\bibitem{r9} V.N. Baier, V.S. Fadin, V. Khoze and E.A. Kuraev, Phys. Rep.
78 (1981) 294; \\
V.M.~Budnev, I.F.~Ginzburg, G.V.~Meledin and V.G.~Serbo,
Phys. Rep. C15 (1975) 183.

\bibitem{gorsh}
V.G.~Gorshkov, Uspekhi Fiz. Nauk 110 (1973) 45.

\bibitem{r10}
R. Barbieri, J.A. Mignaco and E. Remiddi,
Il Nuovo Cimento  11A (1972) 824.

\bibitem{r11}
E.A.~Kuraev, N.P.~Merenkov, V.S.~Fadin,
Sov. J. Nucl. Phys.  45 (1987) 486.

\bibitem{r12}
H.~Cheng and T.T.~Wu,  Expanding Protons: Scattering at High
Energies, London, England, 1986.

\bibitem{r13} N.P.~Merenkov, Sov. J. Nucl. Phys. 48 (1988) 1073.

\bibitem{r14} N.P.~Merenkov, Sov. J. Nucl. Phys. 50 (1989) 469.

\bibitem{r15}
L.N. Lipatov, Sov. J. Nucl. Phys. 20 (1974) 94;\\
G. Altarelli and G. Parisi, Nucl. Phys. B126 (1977) 298;\\
E. A. Kuraev and V. S. Fadin, Sov. J. of Nucl. Phys. 41 (1985) 466;
Preprint INP 84-44, Novosibirsk, 1984; \\
O. Nicrosini and L. Trentadue, Phys. Lett. 196B (1987) 551.

\bibitem{r16}
M. Skrzypek, Acta Phys. Pol. B23 (1993), 135; \\
E.A.~Kuraev, N.P.~Merenkov and V.S.~Fadin,
Sov. J. Nucl. Phys. 47 (1988) 1009. \\
S.~Jadach, M.~Skrzypek, B.F.L.~Ward, Phys. Lett. B257 (1991) p.173.

\bibitem{r17}
%pairs
A.B.~Arbuzov, E.A.~Kuraev, N.P.~Merenkov and L.~Trentadue,
JETP 108 (1995) 1164;
preprint CERN--TH/95--241, JINR E2--95--110, Dubna, 1995.

\bibitem{r18} A.B.~Arbuzov, V.S. Fadin, E.A. Kuraev, L.N. Lipatov,
N.P. Merenkov and L. Trentadue, Report CERN 95-03 (1995) p. 369.


\bibitem{rele}
V.N.~Baier, V.S.~Fadin, V.M.~Katkov,
{\em Emission of relativistic electrons,\/} Moscow, Atomizdat, 1973.

\bibitem{nllbha}
A short write-up of the NLLBHA code can be found in the
CERN Yellow Report CERN-96-01, vol.2. A copy of the program is available,
upon request, from the authors.

\end{thebibliography}
\end{document}